\documentclass[aps, prl, amsmath, amssymb, reprint, floatfix]{revtex4-1}

\usepackage[utf8]{inputenc}
\usepackage[T1]{fontenc}
\usepackage{mathptmx}
\usepackage{etoolbox}
\usepackage{graphicx}
\usepackage{xcolor}
\usepackage{siunitx}
\usepackage[hidelinks]{hyperref}
\usepackage{tabularx}
\usepackage{xspace}
\usepackage{comment}
\usepackage{multirow}
\newcommand{\fig}{Fig.~}
\newcommand{\eq}{Eq.~}
\newcommand{\myRef}{Ref.}

\begin{document}

\title{Neuromorphic spintronics accelerated by an unconventional data-driven Thiele equation approach}

\author{Anatole MOUREAUX}
\affiliation{Institute of Condensed Matter and Nanosciences, Université catholique de Louvain, Place Croix du Sud 1, 1348 Louvain-la-Neuve, Belgium}

\author{Simon DE WERGIFOSSE}
\affiliation{Institute of Condensed Matter and Nanosciences, Université catholique de Louvain, Place Croix du Sud 1, 1348 Louvain-la-Neuve, Belgium}

\author{Chloé CHOPIN}
\affiliation{Institute of Condensed Matter and Nanosciences, Université catholique de Louvain, Place Croix du Sud 1, 1348 Louvain-la-Neuve, Belgium}

\author{Jimmy WEBER}
\affiliation{Institute of Condensed Matter and Nanosciences, Université catholique de Louvain, Place Croix du Sud 1, 1348 Louvain-la-Neuve, Belgium}

\author{Flavio~ABREU~ARAUJO}
\email[]{flavio.abreuaraujo@uclouvain.be}
\homepage[]{https://flavio.be}
\affiliation{Institute of Condensed Matter and Nanosciences, Université catholique de Louvain, Place Croix du Sud 1, 1348 Louvain-la-Neuve, Belgium}

\begin{abstract}
We design a neural network based on a single spin-torque vortex nano-oscillator (STVO) multiplexed in time. The behavior of the STVO is simulated with an improved ultra-fast and quantitative model based on the Thiele equation approach. Different mathematical and numerical adaptations are brought to the model in order to increase the accuracy and the speed of the simulations. We demonstrate the high added value and adaptability of such a neural network through the resolution of three standard machine learning tasks in the framework of reservoir computing. The first one is a task of waveform (sines and squares) classification.
We show the ability of the system to effectively classify waveforms with high accuracy and low root-mean-square error thanks to the intrinsic short-term memory of the device. Given the high throughput of the simulations, two innovative parametric studies on the intensity of the input signal and the level of noise in the system are performed to demonstrate the value of our new models. The efficiency of our system is then tested during a speech recognition task on the TI-$46$ dataset and shows the agreement between the new models and the corresponding experimental measurements. Finally, we use our STVO-based neural network to perform image recognition on the MNIST dataset. State-of-the-art performances are demonstrated, and the interest of using the STVO dynamics as an activation function is highlighted. These results support and facilitate the future development of neuromorphic STVO-based hardware for energy-efficient machine learning.
\end{abstract}

\maketitle

\section{Introduction}

The need for low power and efficient hardware dedicated to machine learning has led to a new type of data processing called neuromorphic computing~\cite{indiveri2011frontiers}.
By taking inspiration from the brain, it tries to overcome the von Neumann bottleneck by proposing artificial neurons or synapses that are highly interconnected within a parallel architecture.
Different systems are under study like photonics~\cite{paquot2010reservoir, paquot2012optoelectronic, larger2012photonic, larger2017high, shastri2021photonics}, memristors~\cite{jo2010nanoscale,du2017reservoir} and spintronics~\cite{grollier2020neuromorphic}.
Among the latter, spin-torque vortex nano-oscillators (STVOs) have already been shown to be choice candidates to implement hardware neurons for machine learning applications. Thanks to their highly non-linear behavior and intrinsic short-term memory, %
as well as a remarkable signal-to-noise ratio despite their nano-metric size, several machine learning tasks such as waveform and speech recognition have been performed successfully with STVO-based neural networks~\cite{torrejon_riou_araujo_tsunegi_khalsa_querlioz_bortolotti_cros_yakushiji_fukushima_2017, riou_torrejon_garitaine_abreu_2019, markovic_leroux_riou_abreu_araujo_torrejon_querlioz_fukushima_yuasa_trastoy_bortolotti_2019}.
Their small size, low power consumption and CMOS-compatibility reinforce their potential interest concerning the development of neuromorphic computing systems~\cite{yogendra2015lowpower, torrejon_riou_araujo_tsunegi_khalsa_querlioz_bortolotti_cros_yakushiji_fukushima_2017}. 

As shown in \myRef~\cite{torrejon_riou_araujo_tsunegi_khalsa_querlioz_bortolotti_cros_yakushiji_fukushima_2017, riou2017neuromorphic, riou2021reservoir}, the recognition success rate is very sensitive to the non-linear profile of the STVO as well as the noise regime. These parameters depend on both, the experimental parameters (external magnetic field and working current intensity for instance) as well as the design of the STVO itself. %
The simulation of such neural networks is hence of prior importance to better understand the underlying phenomena involved in their cognitive properties, and to optimize these systems before the actual fabrication.
Thus, a fast and quantitative model is needed.
Indeed, it would require a huge computational power and an extensive amount of time to study with micromagnetic simulations the dynamics of such oscillators as well as to test several neuromorphic architectures or input parameters that could be optimized for experimental measurements~\cite{mumax3,abreu_araujo_chopin_de_wergifosse_2022}. %
Several solutions are proposed like using a model based on phenomenological non-linear magnetic oscillator theory~\cite{slavin2009nonlinear, abreu_araujo_riou_torrejon_tsunegi_querlioz_yakushiji_fukushima_kubota_yuasa_stiles_2020}, using machine learning to predict the dynamics of STVOs with Neural ODEs~\cite{chen2022forecasting} or using analytical models based on the Thiele equation approach (TEA)~\cite{thiele1973steady, huber1982dynamics} for simulating STVOs.

Abreu Araujo \textit{et al.}~\cite{abreu_araujo_riou_torrejon_tsunegi_querlioz_yakushiji_fukushima_kubota_yuasa_stiles_2020} have already compared experimental results~\cite{torrejon_riou_araujo_tsunegi_khalsa_querlioz_bortolotti_cros_yakushiji_fukushima_2017} and results from STVOs simulated with the non-linear magnetic oscillator theory~\cite{slavin2009nonlinear} for the recognition of spoken digit with reservoir computing.
The parameters needed for the non-linear magnetic oscillator model are extracted experimentally and there is an excellent agreement between experimental and simulated results.
However, the accuracy of the experimental neural network is surprisingly higher than that of the simulated neural network.
On the contrary, the function given by Neural ODEs with the addition of noise allows to predict the experimental results of Torrejon \textit{et al.}~\cite{torrejon_riou_araujo_tsunegi_khalsa_querlioz_bortolotti_cros_yakushiji_fukushima_2017} perfectly with an acceleration factor of $200$ compared to micromagnetic simulations~\cite{chen2022forecasting}.
Still, the function given by the neural network is a black box and one does not have access to the underlying physics of the oscillator. 
As an alternative, TEA models are elegant solutions that have the interest of being based on an analytical description of the underlying physics.
However, they only give quantitative results for the STVO behavior in the resonant regime~\cite{abreu_araujo_chopin_de_wergifosse_2022_splitting} (resp.\ steady-state regime~\cite{guslienko2014nonlinear}) \textit{i.e.} when it does not oscillate (resp.\ when it undergoes stable oscillations).
For the transient regime of STVOs (\textit{i.e.} from the resonant regime to the steady-state regime or vice versa), TEA models are only able to yield qualitative results. Unfortunately, the transient regime is precisely the regime of interest for reservoir computing applications. 
A recent data-driven TEA (DD-TEA) model~\cite{abreu_araujo_chopin_de_wergifosse_2022} has been shown able to describe quantitatively both the steady-state and transient regimes of STVOs. Furthermore, the results were obtained with an acceleration of $6$ orders of magnitude compared to micromagnetic simulations. The resulting increase of the simulations throughput allows to simulate multi-STVOs systems and carry extended parametric studies. This quantitative and ultra-fast model is simply based on TEA and a few micromagnetic simulations.\\
\indent The speed and the accuracy of this DD-TEA model can be further improved as shown by the two analytical models proposed below. %
Indeed, the combination of the DD-TEA model with additional mathematical adjustments allows to solve it fully analytically. The two resulting analytical models described later reach a $9$ orders of magnitude acceleration compared to micromagnetic simulations. These two models are then used to rapidly simulate the dynamics of a STVO. We then design a STVO-based neural network multiplexed through time that we use to solve three machine learning tasks in the framework of reservoir computing~\cite{tanaka2019recent}. The first task is a proof of concept of waveform recognition where sine and square periods have to be distinguished~\cite{paquot2012optoelectronic, riou2017neuromorphic, riou_torrejon_garitaine_abreu_2019,markovic_leroux_riou_abreu_araujo_torrejon_querlioz_fukushima_yuasa_trastoy_bortolotti_2019}. Parametric studies on the accuracy and confidence of the classification are also performed depending on the dc input current intensity and the signal-to-noise ratio (SNR) of the system.
The second task is spoken digits recognition using the TI-$46$ dataset. We compare the results obtained with our two new models with experimental results from Torrejon \textit{et al.}~\cite{torrejon_riou_araujo_tsunegi_khalsa_querlioz_bortolotti_cros_yakushiji_fukushima_2017} and the phenomenological model based on the non-linear magnetic oscillator theory from Abreu Araujo \textit{et al.}~\cite{abreu_araujo_riou_torrejon_tsunegi_querlioz_yakushiji_fukushima_kubota_yuasa_stiles_2020}. The last task is the application of our STVO-based neural network on the MNIST dataset~\cite{deng2012mnist} for written digits recognition. We compare the STVO dynamics with conventional activation functions such as the reLU~\cite{nair2010rectified} and the sigmoid function~\cite{narayan1997generalized} and assess the accuracy of the recognition depending on the number of virtual neurons in the network.  

\section{Methods}\label{section:methods}

The vortex core dynamics can be described by a simple harmonic oscillator equation~\cite{abreu_araujo_chopin_de_wergifosse_2022_splitting} when the pulsation $\omega$ is constant and the transient dynamic factor $\Gamma \rightarrow 0$:
\begin{equation}
    \label{eq:x_harm}
    \left[ {\begin{array}{c}
   X(t) \\
   Y(t) \\
  \end{array} } \right]=
  ||\mathbf{X}||\text{e}^{\Gamma t}
  \left[ {\begin{array}{c}
   \sin{(\omega t)} \\
   -\cos{(\omega t)} \\
  \end{array} } \right]
\end{equation}

where $X(t)$ and $Y(t)$ are the time-dependent Cartesian coordinates of the vortex core. As both $X(t)$ and $Y(t)$ vary quickly in time, \eq(\ref{eq:x_harm}) can be rewritten by considering the reduced vortex core position $s(t) = \sqrt{ X^2(t) + Y^2(t) }/R$ where $R$ is the radius of the magnetic dot.
This gives:
\begin{equation}
    \label{eq:s_harm}
    s(t) = s_\infty \text{e}^{\Gamma t}
\end{equation}
where $s_\infty$ is the final reduced position of the vortex core and $\Gamma$ is a constant related to the transient state of the dynamics.
In reality, $\Gamma$ depends on $s(t)$ and \eq(\ref{eq:x_harm}) admits a general solution that writes:
\begin{equation}
    \label{eq:s_harm_t}
    s(t) = s_\infty \exp{\int_0^t\Gamma(s(t^\prime))\text{d}t^\prime}
\end{equation}

After a few developments not detailed here, \eq(\ref{eq:s_harm_t}) can be expressed as the following ordinary differential equation (ODE):
\begin{equation}
    \label{eq:deriv}
    \dot{s}(t) = \Gamma(s(t))s(t)
\end{equation}
The fully analytical expression of $\Gamma(s)$ as a function of $s$ based on TEA is given in a previous publication~\cite{abreu_araujo_chopin_de_wergifosse_2022_splitting}.
For the sake of simplicity, $\Gamma(s)$ is truncated to the second order:
\begin{equation}
    \label{eq:gamma}
    \Gamma(s) = \alpha + \beta s^2
\end{equation}
Both $\alpha$ and $\beta$ are parameters that depend on the input current density $J$ and are described as follows:
\begin{equation}
    \label{eq:alpha}
    \alpha(J) = a_J J + a
\end{equation}
\begin{equation}
    \label{eq:beta}
    \beta(J) = b_J J + b
\end{equation}
where $a$, $a_J$, $b$, $b_J$ are fully analytically described constants whose value is listed in Table~\ref{tab:constants}.
\begin{table}
\caption{Dynamical constants for the DD-TEA models in the C$+$ chirality}\label{tab:constants}
\begin{ruledtabular}
\begin{tabular}{lrl}
\textbf{Constant} & \multicolumn{2}{c}{\textbf{Value}}\\
\hline
$a_J$ & $6.64$ & Hz cm$^2$A$^{-1}$\\
$b_J$ & $-0.43$ & Hz cm$^2$A$^{-1}$\\
$a$ & $-39.97$ & MHz\\
$b$ & $-25.92$ & MHz
\end{tabular}
\end{ruledtabular}
\end{table}
Finally, by injecting  \eq(\ref{eq:gamma}) in \eq(\ref{eq:deriv}), the following equation appears:

\begin{equation}
    \label{eq:bern}
    \dot{s}(t) = \alpha s(t) + \beta s^3(t)
\end{equation}
\eq(\ref{eq:bern}) is a Bernoulli differential equation of order $3$, which accepts the following solution:
\begin{equation}
    \label{eq:lotea}
    s(t) = \dfrac{s_0}{\sqrt{\left(1+\dfrac{s_0^2}{\alpha/\beta}\right)\exp(-2\alpha t)-\dfrac{s_0^2}{\alpha/\beta}}}
\end{equation}
Where $s_0$ is the initial reduced vortex core position at $t=0$.
The final position of $s$, \textit{i.e.} when $t\rightarrow \infty$, depends on the input current density $J$:
\begin{equation}
    \label{eq:sinf}
    s_\infty(J) = \sqrt{\dfrac{-\alpha(J)}{\beta(J)}}
\end{equation}

Guslienko~\textit{et al.}~\cite{guslienko2014nonlinear} have reported a similar expression of the transient regime obtained with a different method.
The simulations based on this analytical \textit{low-order} truncated model called $s$-LOTEA are two orders of magnitude faster than the previous TEA-based model (DD-TEA~\cite{abreu_araujo_chopin_de_wergifosse_2022}) as no ODE needs to be numerically solved.
The value of $\alpha(J)$ and $\beta(J)$ can be retrieved using the same data-driven approach as for the DD-TEA model~\cite{abreu_araujo_chopin_de_wergifosse_2022}, by fitting the results of a few micromagnetic simulations (performed using mumax$^{3}$~\cite{mumax3}). This leads to a precise description of the transient and steady-state regimes.
This model hence combines physical foundations and the precision of a data-driven method.

The second model is an extension of the first one, allowing to capture any additional non-linearity in the dynamics of the vortex core that would not be accounted for in \eq(\ref{eq:lotea}) due to the truncation to the second order of the $\Gamma(s)$ parameter. To do so, a purely mathematical adjustment is brought to the $s$-LOTEA model. Indeed, Bernoulli differential equations such as \eq(\ref{eq:bern}) can involve any real power. Hence, \eq(\ref{eq:lotea}) can be generalized to the order $n$ to write \eq(\ref{eq:hptea}), where $n(J)$ is a fifth-order polynomial of the current density $J$ injected into the STVO, whose coefficients have been determined by fitting to micromagnetic simulations results. This model was hence called the \textit{s-analytical High-Precision Thiele equation approach} model ($s$-HPTEA), as it takes into account the whole complexity of the vortex core dynamics.
\begin{equation}
    \label{eq:hptea}
    s(t) = \dfrac{s_0}{\sqrt[n]{\left(1+\dfrac{s_0^n}{\alpha/\beta}\right)\exp(-n\alpha t)-\dfrac{s_0^n}{\alpha/\beta}}}
\end{equation}

These two models, on top of being extremely fast to compute compared to micromagnetic simulations, yield physically accurate results thanks to the combination with micromagnetic simulations data. The good agreement between micromagnetic simulations and the DD-TEA framework has already been thoroughly presented by Abreu Araujo \textit{et al.}~\cite{abreu_araujo_chopin_de_wergifosse_2022}.

Our models were then used to simulate the output of a STVO for a given input signal. A STVO-based neural reservoir~\cite{paquot2010reservoir, larger2012photonic, paquot2012optoelectronic, larger2017high, schrauwen2007overview} was designed. By using time multiplexing, this recurrent network was emulated with only one STVO~\cite{torrejon_riou_araujo_tsunegi_khalsa_querlioz_bortolotti_cros_yakushiji_fukushima_2017}. Instead of connecting several different STVOs in space to compose it, the unique STVO successively played the role of the different virtual neurons connected through time, hence allowing to simplify the setup. 
The training and inference processes are thoroughly described in \myRef~\cite{riou_torrejon_garitaine_abreu_2019, abreu_araujo_riou_torrejon_tsunegi_querlioz_yakushiji_fukushima_kubota_yuasa_stiles_2020}, and are similar to the ones presented in \myRef~\cite{furuta2018macromagnetic,larger2012photonic,larger2017high,paquot2010reservoir, paquot2012optoelectronic, tanaka2019recent}.

\section{Benchmarking}
\subsection{Waveform recognition}

\begin{figure}
    \centering%
    \includegraphics[width=0.48\textwidth]{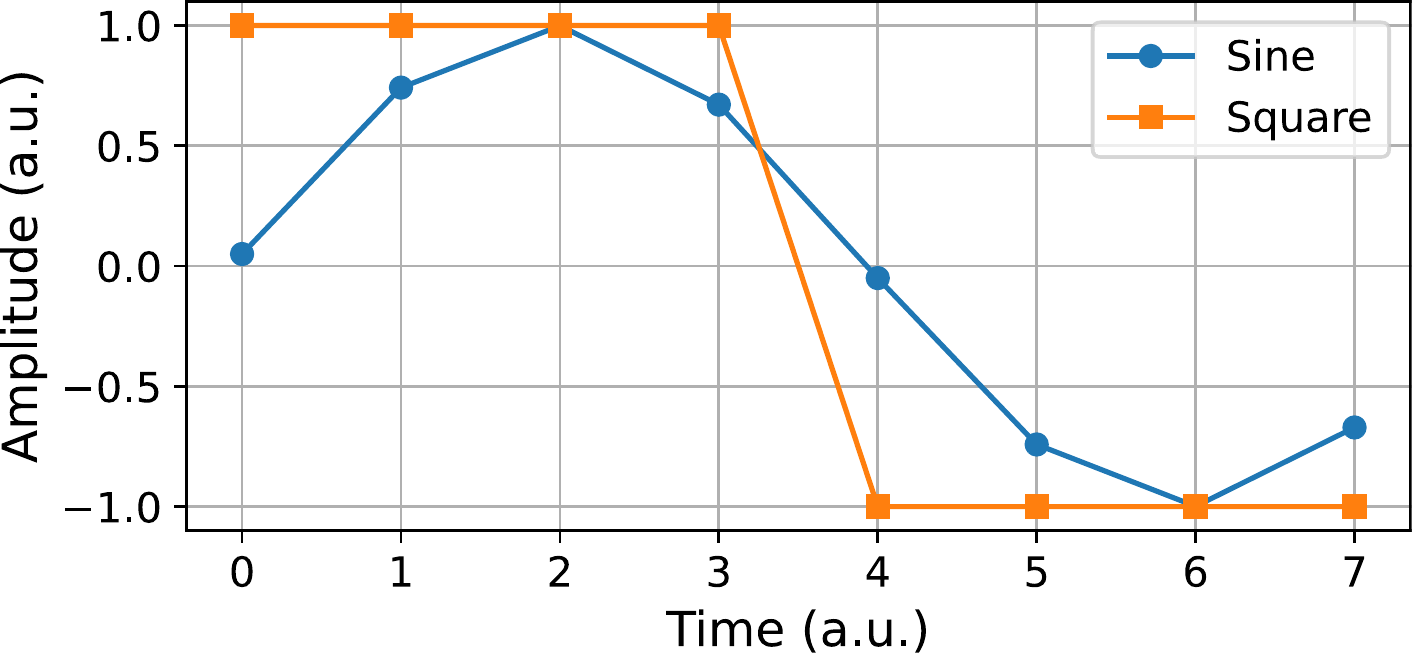}
    \caption{Sine and square input signals used for benchmark.}\label{fig:signals}
\end{figure}

We first performed classification of sine and square periods~\cite{paquot2012optoelectronic, riou2017neuromorphic, riou_torrejon_garitaine_abreu_2019,markovic_leroux_riou_abreu_araujo_torrejon_querlioz_fukushima_yuasa_trastoy_bortolotti_2019} composed of $8$ samples (\fig\ref{fig:signals}) using a time-multiplexed reservoir of $24$ neurons as prescribed in \myRef~\cite{riou_torrejon_garitaine_abreu_2019}. 
The intrinsic dynamics of the STVO allowing to generate the output signal was modeled using the two models from \eq(\ref{eq:lotea}) and \eq(\ref{eq:hptea}) successively, allowing to compare their respective performance. 

To train the network, a database of $80$ sine periods and $80$ square periods arranged in an alternate fashion was used. 
For the testing, a similar yet randomly arranged database was preferred. 

The classification result yielded by the neural network was constructed in two different ways, as presented in \myRef~\cite{abreu_araujo_riou_torrejon_tsunegi_querlioz_yakushiji_fukushima_kubota_yuasa_stiles_2020}. 
The $\tau$-Wise (TW) approach consists in treating each of the $8$ samples of a given period individually. The accuracy is thus a value between $0\%$ if none of the period samples is correctly recognized, and $100\%$ if the $8$ samples are correctly recognized. This allows to assess the performance of the neural network at recognizing sometimes very small partial inputs of data (in this case one eighth of a period). 
The Winner-Takes-All (WTA) method involves averaging the classification of the $8$ samples of a period before inferring the classification of the total period. This leads to an accuracy of $100\%$ if the final value corresponds to the expected target, or $0\%$ if not. This technique allows to absorb any small inaccuracies that may occur during the recognition of a period by considering it globally, hence leading to a slightly better accuracy than with the TW approach.

The geometry of a known experimental STVO was fully mimicked during the simulations. The diameter of the STVO was fixed to $d=200$ nm, while its dc resistance $R_\text{osc}$ was set to $140.6$ $\Omega$, accordingly to the values presented in \myRef~\cite{torrejon_riou_araujo_tsunegi_khalsa_querlioz_bortolotti_cros_yakushiji_fukushima_2017,riou_torrejon_garitaine_abreu_2019,abreu_araujo_riou_torrejon_tsunegi_querlioz_yakushiji_fukushima_kubota_yuasa_stiles_2020}. 
The chirality of the STVO~\cite{abreu_araujo_chopin_de_wergifosse_2022_splitting} was fixed to $+1$ during all the simulations to ensure the consistency of the results. However, all the calculations and results are extendable to the other chirality ($-1$) by using the corresponding constants~\cite{abreu_araujo_chopin_de_wergifosse_2022_splitting}.

\begin{figure}
    \centering%
    \includegraphics[width=0.48\textwidth]{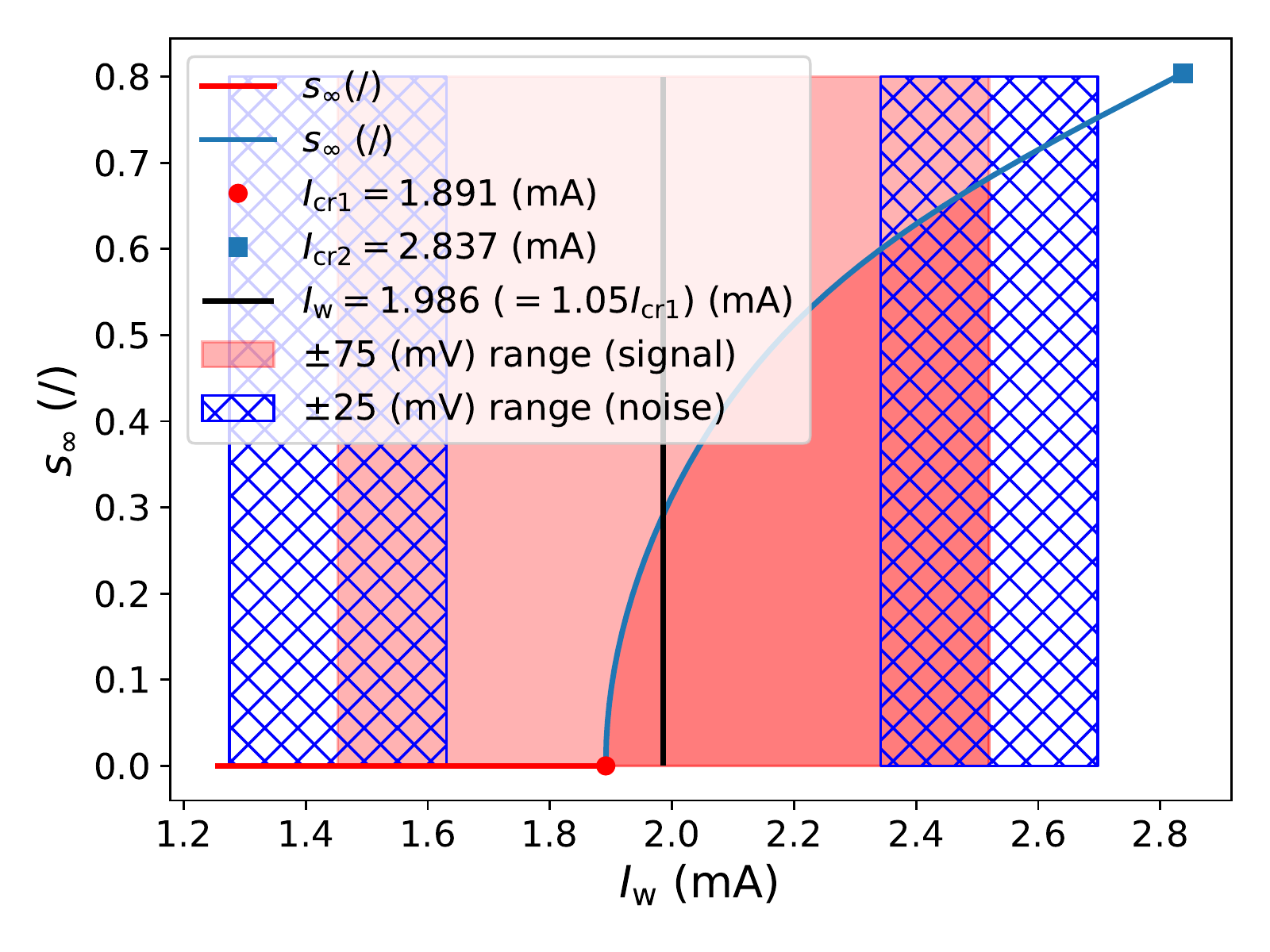}
    \caption{Example of the current map and the range sounded by the dynamics of the oscillator. The black line corresponds to the working current intensity $I_\text{w}$. The red range corresponds to the $150$ mV peak-to-peak voltage amplitude $\Delta V$ of the input signal $V_\text{signal}$, and the blue hatched ranges correspond to an additional peak-to-peak noise amplitude of $\Delta V_\text{noise} = 50$ mV. The curve is the steady-state reduced position of the vortex core $s_\infty$ reached for each input current intensity.}\label{fig:map}
\end{figure}

The dc bias current intensity, or working current intensity $I_\text{w}$ is used to trigger the gyrotropic motion of the vortex core. The input signal $V_\text{signal}$ is added to it before entering the STVO as described in \eq (\ref{eq:input}). The value of $I_\text{w}$ is important as it defines the regime in which the STVO will operate, and hence influences the way the data is processed by the STVO\@. It is represented by the black line in \fig\ref{fig:map}. In our base case, $I_\text{w}$ was set to $1.986 = 1.05\times I_\text{cr1}$ mA, with $I_\text{cr1}$ the first critical current intensity under which the STVO lies in the resonant state~\cite{abreu_araujo_chopin_de_wergifosse_2022_splitting}. This was done to optimize the use of the transient STVO dynamics in the data treatment. However this value can actually be swept across a range of values in order to assess the influence of $I_\text{w}$ on the results of the neural network during the classification. 
The power of the input signal can be written as in \eq(\ref{eq:ps1}) and \eq(\ref{eq:ps2}).
\begin{align}
P_\text{signal} &= R_\text{osc}I_\text{w}^2&(\text{W})
\label{eq:ps1}
\\
&= 10\times \text{log}_{10}\left(R_\text{osc}I_\text{w}^2\right)&(\text{dB})
\label{eq:ps2}
\end{align}

The peak-to-peak amplitude of the input signal $V_\text{signal}$ was scaled up to $\Delta V = 150$ mV. It is represented by the red interval around the $I_\text{w}$ line in \fig\ref{fig:map}. 

The sampling rate of the input signal was characterized with a time constant $D_\text{t}$ of $50$ ns. This parameter must also be chosen wisely as the transient state of the vortex core dynamics is the main contribution to the non-linear treatment of the data. Hence, a too small $D_\text{t}$ will not allow to sufficiently leverage the transient state of the dynamics, while a too long $D_\text{t}$ will lead to the saturation of the regime into the steady-state, hence decreasing the efficiency of the data treatment. 

White noise from various sources can be found in the experimental system under the form of thermal, electrical or magnetic noise. To consider the noise from all these sources as a single resultant quantity, Gaussian white noise was added to the input signal. We used a normal distribution defined with a mean amplitude of $0$ mV and a standard deviation $\sigma$ chosen to yield a predefined peak-to-peak amplitude $\Delta V_\text{noise, p-2-p}$ during $99.7\%$ of the time accordingly to the $6$-sigma rule and \eq(\ref{eq:sigma}). In our base case, $V_\text{noise, p-2-p}$ was set to $50$ mV.
\begin{equation}
\sigma \simeq \dfrac{\Delta V_\text{noise, p-2-p}}{6\sqrt{R_\text{osc}}}\quad (\text{W}^{1/2})
    \label{eq:sigma}
\end{equation} 
The white noise can then be expressed as \eq(\ref{eq:noiseV}) and the resulting input signal can be written as \eq(\ref{eq:input}). 
\begin{equation}
    \label{eq:noiseV}
    V_\text{noise} = \text{normal}(\mu=0, \sigma)
\end{equation}
\begin{equation}
    \label{eq:input}
    I_\text{input} = I_\text{w}+\dfrac{V_\text{signal}+V_\text{noise}}{R_\text{osc}}\quad (\text{mA})
\end{equation}

The power of the noise can then be expressed as \eq(\ref{eq:pn1}), \eq(\ref{eq:pn2}) and \eq(\ref{eq:pn3}). In the base case (\textit{i.e.} when $\Delta V_\text{noise, p-2-p} = 50$ mV), it is equal to $-33.1$ dBm. Note that in practice, there is no noise in the input waveforms and all the noise originates from the STVO\@. However, as the analytical models do not model this internal noise, it had to be artificially added to the input signal. 
\begin{align}
P_\text{n} &= \sigma^2&(\text{W})
\label{eq:pn1}
\\
&= 10\times \text{log}_{10}\left(\sigma^2\right)&(\text{dB})
\label{eq:pn2}
\\
&= 20\times \text{log}_{10}\left(\sigma\right)+30&(\text{dBm})
\label{eq:pn3}
\end{align}

The resulting SNR can be computed using \eq(\ref{eq:snr_iw1}), \eq(\ref{eq:snr_iw2}) and \eq(\ref{eq:snr_iw3}). In the base case related to \fig\ref{fig:map} (\textit{i.e.} when $I_\text{w} = 1.05\times I_\text{cr1}$ and $\Delta V_\text{noise, p-2-p} = 50$ mV), it is equal to $30.5$ dB. The level of noise can also be swept on a defined range to assess its influence on the accuracy of the recognition. 
\begin{align}
\text{SNR}&=\dfrac{R_\text{osc}I_\text{w}^2}{\sigma^2} = \dfrac{36\times R_\text{osc}^2I_\text{w}^2}{\Delta V_\text{noise, p-2-p}^2}&(/)
\label{eq:snr_iw1}
\\
&=10\times\text{log}_{10}\left(\dfrac{36\times R_\text{osc}^2I_\text{w}^2}{\Delta V_\text{noise, p-2-p}^2}\right)&(\text{dB})
\label{eq:snr_iw2}\\
&=20\times \text{log}_{10}\left(\dfrac{6\times R_\text{osc}I_\text{w}}{\Delta V_\text{noise, p-2-p}}\right)&(\text{dB})
\label{eq:snr_iw3}
\end{align}


The results of the recognition were analyzed with two indicators: the accuracy and the root-mean-square of the error (RMSE) between the expected and actual outputs $T_\sigma$ and $\hat{T}_\sigma$ (\eq\ref{eq:rms}). The latter indicator was used as the accuracy was close to $100\%$ in the large majority of the cases due to the low complexity of the task. High RMSE values indicate a less confident recognition even if the classification is correct. These indicators were averaged over the $80$ signals of the testing database.

\begin{equation}
    \label{eq:rms}
    \text{RMSE} = \sqrt{\overline{{\left(T_\sigma-\hat{T}_\sigma\right)}^2}}
\end{equation}

The comparison between the results of the $s$-LOTEA and the $s$-HPTEA models during the benchmark task, averaged over $2000$ simulations, is presented in Table~\ref{table:lotea_hptea}. Concerning the performance at the benchmark task, one can observe that in all of the cases but one, the $s$-HPTEA model yields slightly better results than the $s$-LOTEA model. %
The averaging of the results allows to state that these differences are unlikely to be randomly due to the simulated noise but rather to the models themselves. More specifically, this is due to the better description of the complexity of the vortex core dynamics in the $s$-HPTEA model. The $s$-HPTEA model, in addition to
being the most physically accurate so far, thus also yields the best results when performing the benchmark task. This highlights the useful role of the STVO dynamics complexity into their performance as hardware neurons. In a more general way, it can be seen that the accuracy is extremely close to that of a perfect recognition, while needing orders of magnitude less simulation time than micromagnetic simulations. This is made possible thanks to the consideration of past inputs in our new models, which represents the intrinsic short-term memory of STVOs. This memory helps to discriminate between the different classes of waveforms even if some samples are identical in the signals (such as the ones at $t=2$ and $t=6$ in \fig\ref{fig:signals})~\cite{riou_torrejon_garitaine_abreu_2019}.
\begin{table}
\centering%
\caption{Results of the benchmark task for the $s$-LOTEA and the $s$-HPTEA models averaged over $2000$ simulations using the parameters related to \fig\ref{fig:map} (base case). The bold values correspond to the best performance (\textit{i.e.} highest accuracy or lowest RMSE) between the two models.}
\begin{ruledtabular}
\begin{tabular}{llcr}
&\textbf{Metrics}&\textbf{$s$-LOTEA}&\textbf{$s$-HPTEA}\\
\hline
\multirow{4}{*}{\textbf{Training}}&Accuracy (WTA)&$99.99\%$&${\bf 100\%}$\\
&Accuracy (TW)&$99.77\%$&${\bf 99.94\%}$\\
&RMSE (WTA)&$0.235$&${\bf 0.198}$\\
&RMSE (TW)&$0.350$&${\bf 0.300}$\\
\hline
\multirow{4}{*}{\textbf{Testing}}&Accuracy (WTA)&${\bf 99.82\%}$&$99.73\%$\\
&Accuracy (TW)&$99.26\%$&${\bf 99.39\%}$\\
&RMSE (WTA)&$0.278$&${\bf 0.245}$\\
&RMSE (TW)&$0.402$&${\bf 0.348}$
\end{tabular}
\end{ruledtabular}\label{table:lotea_hptea}
\end{table}

\subsubsection{Influence of the working current intensity}
Due to the high throughput of the simulations allowed by the new models, two parametric studies have been performed. The first one consists in a sweep of the working current intensity $I_\text{w}$ to investigate the influence of the operating regime and the corresponding non-linearity on the treatment of the data. The working current intensity was swept from $1.001\times I_\text{cr1}$ to the maximum allowed working current intensity $I_\text{w,max}$ defined in \eq(\ref{eq:iwmax}), ensuring that the total injected current intensity was not exceeding the second critical current intensity $I_\text{cr2}$~\cite{abreu_araujo_chopin_de_wergifosse_2022_splitting} in $99.7\%$ of the cases. $\Delta V_\text{noise, p-2-p}$ was set to $50$ mV.
\begin{equation}
    I_\text{w,max} = I_\text{cr2}-\dfrac{\Delta V+\Delta V_\text{noise, p-2-p}}{2R_\text{osc}}\quad (\text{mA})
    \label{eq:iwmax}
\end{equation}
The other operating parameters are listed in Table~\ref{table:Iw_sweep}. For each value in the range swept, the accuracy and the RMSE were computed for both TW and WTA approaches, and so for both $s$-LOTEA and $s$-HPTEA models. To avoid the random fluctuations introduced with the background noise, the results were averaged over $200$ simulations for each value of the sweep range. Note that the SNR also increases to a lesser extent during the sweep of $I_\text{w}$. Indeed, as $I_\text{w}$ was swept from $1.001\times I_\text{cr1}$ to $I_\text{w,max}$, the SNR was swept between $30.1$ and $33.6$ dB accordingly to \eq(\ref{eq:snr_iw3}).

\begin{figure}
    \centering%
    \includegraphics[width=.48\textwidth]{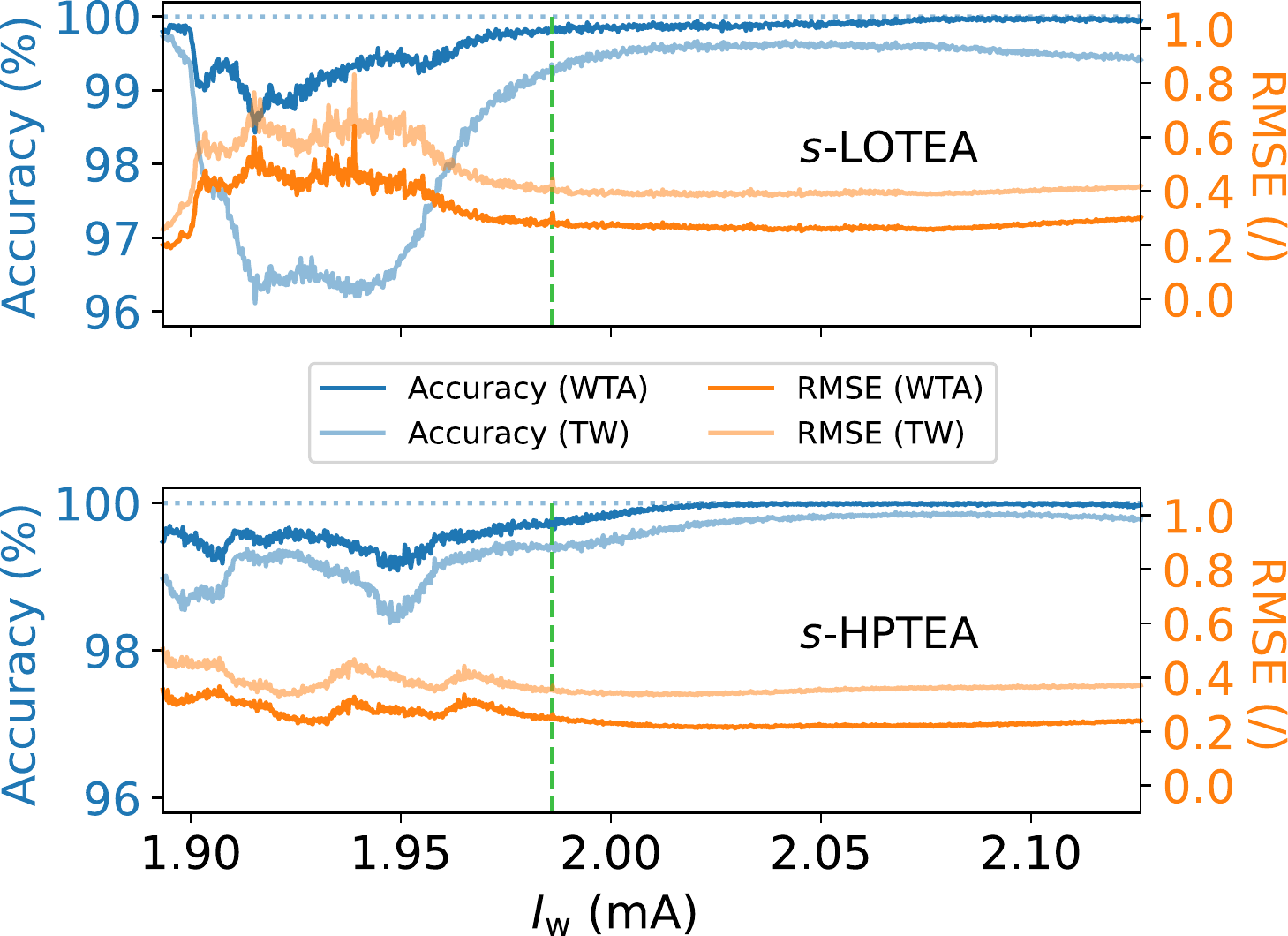}
    \caption{Test accuracy and RMSE of the neural network emulated by a STVO, simulated using the $s$-LOTEA model (top) and the $s$-HPTEA model (bottom) during the parametric sweep of the working current intensity, averaged over $200$ simulations. The green dashed vertical lines represent the working current intensity used in the base case ($1.986$ mA).}\label{fig:iw}
\end{figure}

\begin{figure}
    \centering%
    \includegraphics[width=0.48\textwidth]{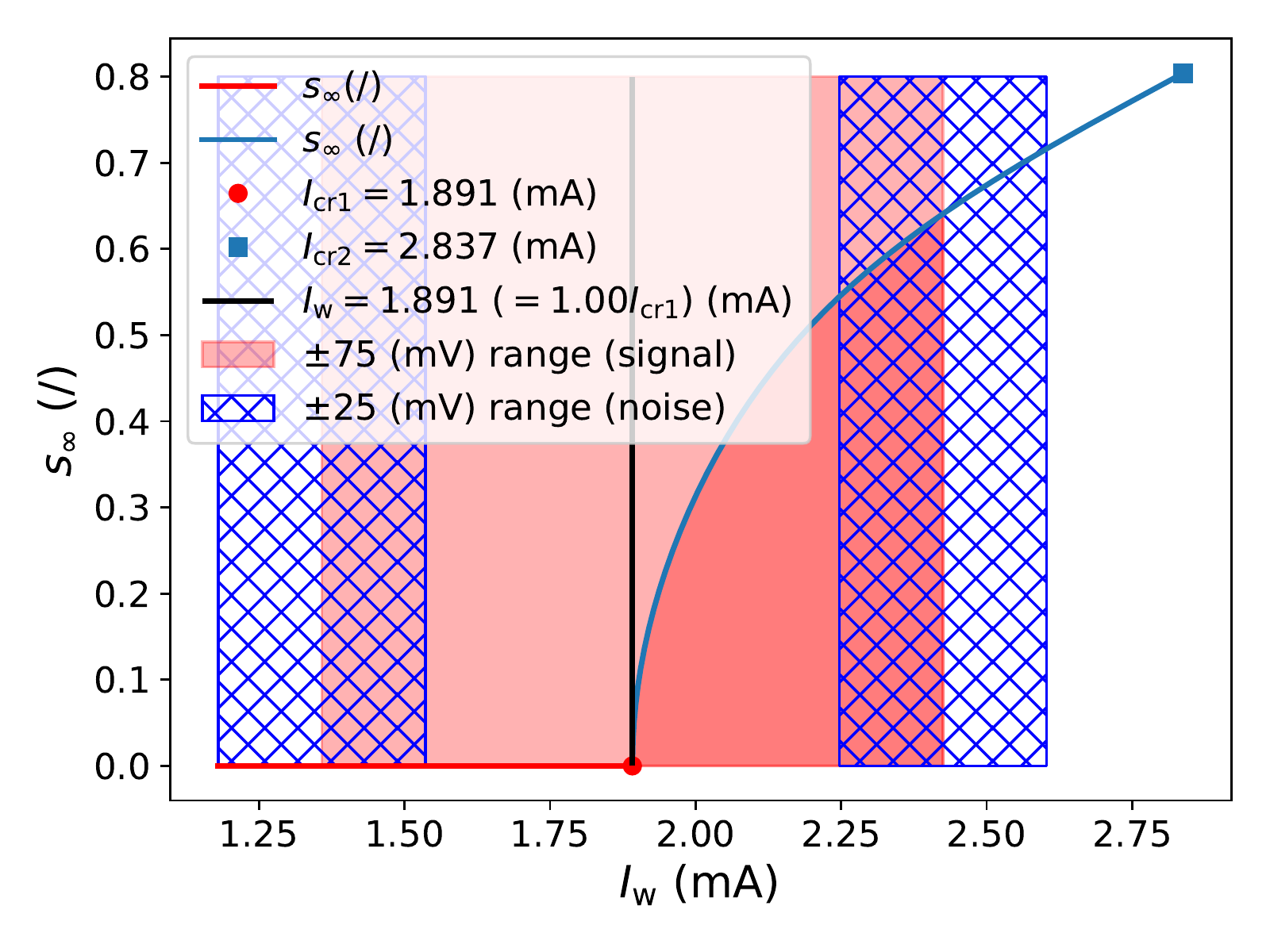}
    \caption{Current map and range sounded by the dynamics of the oscillator when $I_\text{w} = I_\text{cr1}$. The signal has $50\%$ chance to lie below $I_\text{cr1}$ and to prevent the triggering of the STVO dynamics.}\label{fig:map2}
\end{figure}

The results obtained for the parametric sweep of the working current intensity with the two models are displayed in \fig\ref{fig:iw}. It can be seen that for higher working current intensities, the performance of the neural network gets progressively better. Indeed, the accuracy reaches an upper asymptote at $100\%$ and the RMSE decreases monotonically to a lower asymptotic value. This observation is valid for both models. It means that the regime in which the STVO operates when submitted to higher bias current intensities is beneficial for the recognition. %
This is due to the behavior of the STVO being less linear at higher current intensities, allowing a better quality of data processing~\cite{abreu_araujo_riou_torrejon_tsunegi_querlioz_yakushiji_fukushima_kubota_yuasa_stiles_2020}. When $I_\text{w}$ decreases, the probability of the signal $I_\text{input}$ to lie below the first critical current intensity $I_\text{cr1}$ increases. Hence, the signal has a lower probability of triggering the vortex core oscillations required for the effective treatment of the data. For example, if $I_\text{w} = I_\text{cr1}$ as in \fig\ref{fig:map2}, the signal has only $50\%$ chance to lie above $I_\text{cr1}$ and to induce STVO oscillations. This results in a decrease of the cognitive performance that can be seen at the leftmost end of \fig\ref{fig:iw}.

Mathematically, this can be interpreted by considering the $s_\infty$ curve represented in red and blue in \fig\ref{fig:map} and \fig\ref{fig:map2}. Indeed, under $I_\text{cr1}$ the vortex core is in the resonant state and its final reduced position $s_\infty$ is equal to $0$. As $s(t)$ is the effective activation function that allows the treatment of the input data, any data point lying under $I_\text{cr1}$ is mapped to $0$, and cannot be made good use of to train the network. %
If all the signal were to lie below $I_\text{cr1}$, it would not able to trigger the oscillations of the vortex core and the input data would be linearly mapped to $0$. An accuracy corresponding to random choice, \textit{i.e.} $50\%$, could hence be expected. This is verified is Table~\ref{tab:linear}. 

\begin{table}
\centering%
\caption{Results of the benchmark task for the $s$-LOTEA and the $s$-HPTEA models averaged over $2000$ simulations when the input signal completely lies below $I_\text{cr1}$.}
\begin{ruledtabular}
\begin{tabular}{llcr}
&\textbf{Metrics}&\textbf{$s$-LOTEA}&\textbf{$s$-HPTEA}\\
\hline
\multirow{4}{*}{\textbf{Training}}&Accuracy (WTA)&$50.00\%$&${50.00\%}$\\
&Accuracy (TW)&$50.00\%$&${50.00\%}$\\
&RMSE (WTA)&$1.000$&${1.000}$\\
&RMSE (TW)&$1.000$&${1.000}$\\
\hline
\multirow{4}{*}{\textbf{Testing}}&Accuracy (WTA)&${50.00\%}$&$50.00\%$\\
&Accuracy (TW)&$50.00\%$&${50.00\%}$\\
&RMSE (WTA)&$1.000$&${1.000}$\\
&RMSE (TW)&$1.000$&${1.000}$
\end{tabular}
\end{ruledtabular}\label{tab:linear}
\end{table}

More generally, it can be noticed that the WTA approach systematically yields better results (\textit{i.e.} higher accuracy and lower RMSE) than the TW approach as explained before. Furthermore, it can also be seen that the asymptotic values obtained with the $s$-HPTEA model are always better than that obtained with the $s$-LOTEA model. All these observations show an undoubted agreement with the results presented in the lower half of Table~\ref{table:lotea_hptea}, highlighting the consistency of the simulations. 

The existence of an optimal input current intensity is however suspected. As a matter of fact as $I_\text{w}$ increases, the probability of the noise to contribute to the STVO dynamics also increases. %
This noise has a detrimental influence on the vortex core dynamics and hinder the network performance as it will be shown later. However, this decrease of the performance at high input current intensities is relatively small, due to $I_\text{w}$ being bounded by $I_\text{w,max}$ (see \eq(\ref{eq:iwmax})).

This kind of study demonstrates the high value of our new analytical models. Those can be used to guide the design of an experimental system, by specifying the optimal working current intensity required to reach a given accuracy for a given task. In our case, $I_\text{w}$ should be about $2.05$ mA in order to reach an accuracy of $99.99\%$ when considering the $s$-HPTEA model.

\subsubsection{Influence of the noise}
The second parametric study consists in a sweep of the signal-to-noise ratio (SNR) at a given working current intensity. For a given value of the SNR (in dB), Gaussian white noise of the corresponding amplitude was added to the signal (as the blue hatched areas in \fig\ref{fig:map}). A SNR of $0$ dB corresponds to the case where the power of the input signal is equal to that of the added white noise, while positive (resp.~negative) values correspond to the case were the power of the signal is higher (resp.~lower) than that of the added noise. %
The SNR was swept from $-20$ dB to $100$ dB, and the working current intensity was set to $I_\text{w} = 1.05\times I_\text{cr1} = 1.986$ mA. The other operating parameters are listed in Table~\ref{table:Iw_sweep}. The same metrics as in the first parametric study (\textit{i.e.} the accuracy and the RMSE) were computed and averaged over $200$ simulations. The case where SNR $\leq0$ dB is obviously not likely to happen in practice as it would mean that one expects to successfully classify random signals, but was nevertheless investigated for reasons detailed further.  
\begin{table}
\centering%
\caption{Input parameters for the parametric sweeps.}
\begin{ruledtabular}
\begin{tabular}{lcrl}
\textbf{Parameter}&\textbf{Notation}&\multicolumn{2}{c}{\textbf{Value}}\\
\hline
DC resistance&$R_\text{osc}$&$140.6$ & $\Omega$\\
Chirality&$C$&$+1$\\
Sampling rate&$D_\text{t}$&$50$ & ns\\
Amplitude of the input signal&$\Delta V$&$150$ & mV\\
STVO diameter&$d$&$200$ & nm
\end{tabular}
\end{ruledtabular}\label{table:Iw_sweep}
\end{table}

\begin{figure}
    \centering%
    \includegraphics[width=.48\textwidth]{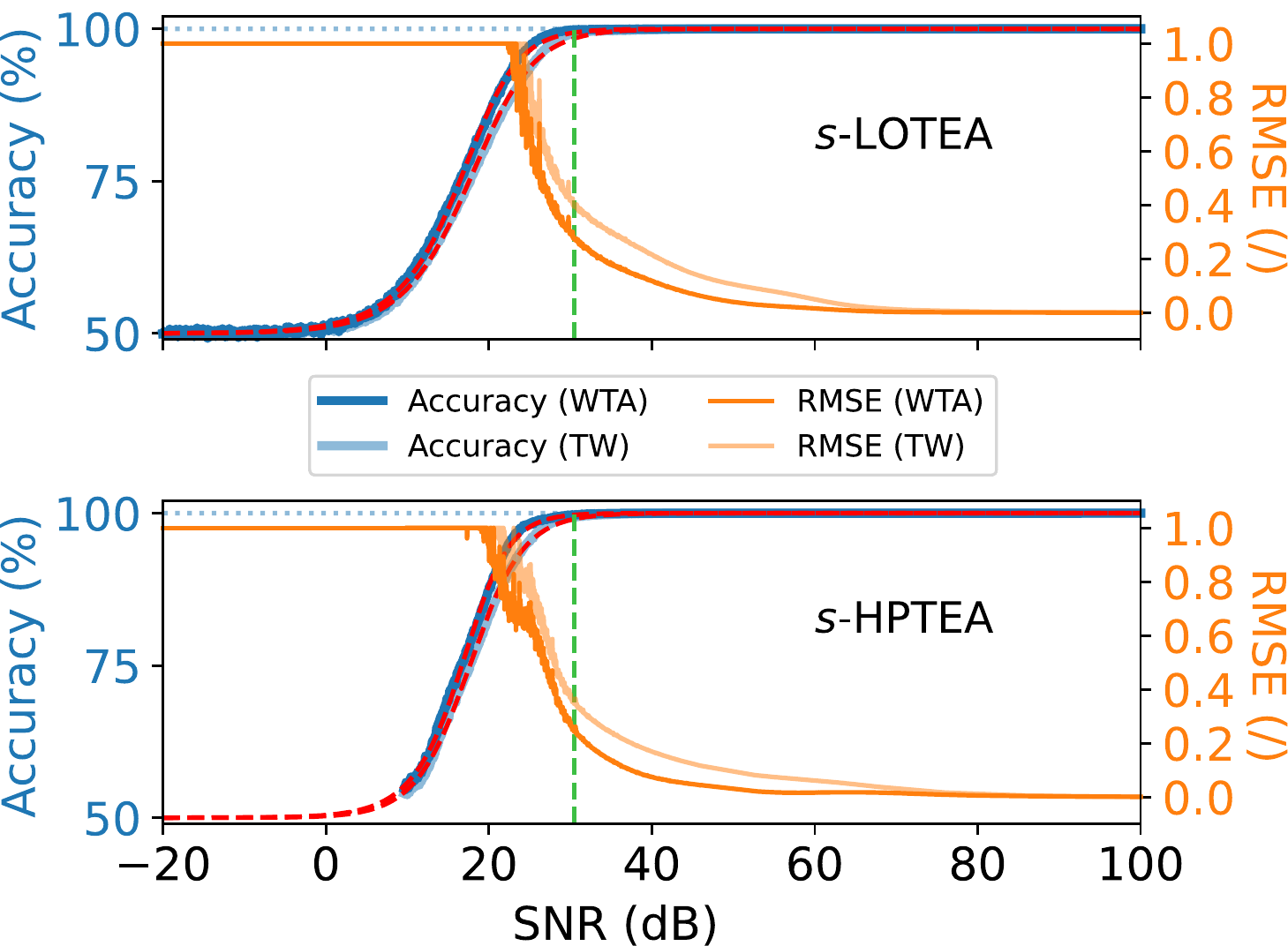}
    \caption{Test accuracy and RMSE of the neural network emulated by a STVO, simulated using the $s$-LOTEA model (top) and the $s$-HPTEA model (bottom) during the parametric sweep of the system SNR, averaged over $200$ simulations. The RMSE values were truncated to $1.00$ due to high RMSE for negative SNRs. The red dashed lines represent the piece-wise generalized logistic approximations from \eq(\ref{eq:genlog}) and Table~\ref{tab:genlogs_slotea}. The green dashed vertical lines represent the SNR simulated in the base case ($30.5$ dB).}\label{fig:snr}
\end{figure}

The results of the parametric study on the SNR are displayed in \fig\ref{fig:snr}. It can be noticed that a sigmoid-like curve is obtained when plotting the accuracy with respect to the SNR\@. For positive SNRs, the accuracy reaches $100\%$ due to the progressively less noisy system. As the SNR decreases, the accuracy drops to $50\%$. %
This accuracy corresponds to random choice between the two categories (sine and square). This is due to the noise decreasing the quality of the STVO dynamics down to the point where no usable features can be found in the output signals and leveraged by the neural network for the classification. The resulting curves were fitted using a generalized logistic approximation such as \eq(\ref{eq:genlog}), whose coefficients and maximum relative errors are presented in Table~\ref{tab:genlogs_slotea}.
\begin{equation}
\label{eq:genlog}
    \text{ACC}(\text{SNR}) = 50+\dfrac{50}{{\left(1+Q\exp(-B\text{ SNR})\right)}^{1/\nu}}\%
\end{equation}
\begin{table}
\caption{Coefficients and maximum relative error of the generalized logistic approximations of the accuracy reached with respect to the SNR (see \eq(\ref{eq:genlog}) and \fig\ref{fig:snr}).}\label{tab:genlogs_slotea}
\begin{ruledtabular}
\begin{tabular}{llrlr}
 & \multicolumn{2}{c}{\textbf{$s$-LOTEA}}& \multicolumn{2}{c}{\textbf{$s$-HPTEA}}\\
 & WTA & TW & WTA & TW\\
\hline
$Q$ & $553$ & $214$ & $1180$ & $259$\\
$B$ & $0.33$ & $0.28$ & $0.39$ & $0.31$\\
$\nu$ & $1.74$ & $1.39$ & $1.49$ & $1.07$\\
$\max (\text{RE})$ & $2.9\%$ & $2.4\%$ & $2.7\%$ & $2.6\%$
\end{tabular}
\end{ruledtabular}
\end{table}

These relations, which would not have been possible to express accurately without the new models, could help to estimate precious information about future physical prototypes. For example the expected accuracy for this recognition task can be retrieved once the value of the SNR of the system has been estimated. Alternatively, these relations could also be used to know the minimum SNR value required to reach a given accuracy level. %
For example, by considering the $s$-HPTEA model, it is possible to state that the SNR must be at least $22.77$ dB to reach an accuracy of $95\%$ with the WTA approach. Considering a working current intensity of $1.986$ mA (as in the base case), this corresponds to a noise whose average peak-to-peak amplitude is about $122$ mV.

The interpretation of the RMSE plots is somewhat less trivial. Under a SNR of about $20$ dB, the RMSE explodes and has to be truncated to $1.00$. Indeed, as the random fluctuations introduced with the noise become more and more predominant for lower SNRs, the resulting input signal becomes less and less similar to the bounded signals that were used to train the network (\fig\ref{fig:signals}). The occurrence of these unexpected high-amplitude values in the network leads to a steep increase of the RMSE\@. As the SNR increases above $20$ dB, the progressively cleaner dynamics leads to a monotonous decrease of the RMSE as expected.

One can observe that there are no results for the accuracy under $10$ dB with the $s$-HPTEA model due to the simulations crashing repeatedly. It is suspected that the $s$-HPTEA model is more sensitive to noise due to its inherent higher-order dynamics.

As previously said, the SNR also increases from $30.1$ dB to $33.6$ dB when $I_\text{w}$ increases. However, the improvement of the performance observed during the parametric sweep of $I_\text{w}$ is much more due to the more complex STVO dynamics at higher input current intensities than to the corresponding increase of the SNR. Indeed, the accuracy gained at higher input current intensities showed in \fig\ref{fig:iw} is higher than the accuracy gained between $30.1$ dB and $33.6$ dB in \fig\ref{fig:snr}.

\subsection{Speech recognition}
The second task to which our new models were applied was the same speech recognition task as in \myRef~\cite{torrejon_riou_araujo_tsunegi_khalsa_querlioz_bortolotti_cros_yakushiji_fukushima_2017, abreu_araujo_riou_torrejon_tsunegi_querlioz_yakushiji_fukushima_kubota_yuasa_stiles_2020} on the TI-$46$ dataset. This was meant to compare the simulated results with pre-existing experimental measurements. 

The SNR of the experimental STVO used in  \myRef~\cite{abreu_araujo_riou_torrejon_tsunegi_querlioz_yakushiji_fukushima_kubota_yuasa_stiles_2020} was estimated. The noise in the input signals was considered negligible and the estimator $\widehat{\text{SNR}}$ was computed as in \eq(\ref{eq:snr_exp}), \eq(\ref{eq:snr_exp2}) and \eq(\ref{eq:snr_exp3}).
\begin{align}
    \label{eq:snr_exp}
    \widehat{\text{SNR}} &= \dfrac{\widehat{P}_\text{signal}}{\widehat{P}_\text{noise}}\\
    \label{eq:snr_exp2}
    &= \dfrac{\widehat{V}_\text{RMS, signal}^2}{\widehat{V}_\text{RMS, noise}^2}\\
    \label{eq:snr_exp3}
    &=\dfrac{\overline{\left(\widehat{V}_\text{signal}^2\right)}}{\overline{\left(\widehat{V}_\text{noise}^2\right)}}
\end{align}
$\widehat{V}_\text{signal}$ and $\widehat{V}_\text{noise}$ are estimations of the amplitude of the signal and the noise retrieved using the amplitude of the raw STVO output signal $V_\text{AVG1}$ and the average of the STVO output signal over $16$ measurements $V_\text{AVG16}$ (see \eq(\ref{eq:sig_estimation}) and \eq(\ref{eq:noise_estimation})). 
\begin{equation}
    \label{eq:sig_estimation}
    \widehat{V}_\text{signal} \approx V_\text{AVG16}
\end{equation}
\begin{equation}
    \label{eq:noise_estimation}
    \widehat{V}_\text{noise} \approx V_\text{AVG1} - V_\text{AVG16}
\end{equation}
The SNRs values obtained for the experimental measurements performed using the MFCC, cochleagram and Spectro HP filters are presented in Table~\ref{tab:snrs}. These positive values induce that one can expect to observe an improvement of the classification metrics (accuracy and RMSE) between the raw $V_\text{AVG1}$ output signals and the less noisy $V_\text{AVG16}$ output signals.

The task described in \myRef~\cite{abreu_araujo_riou_torrejon_tsunegi_querlioz_yakushiji_fukushima_kubota_yuasa_stiles_2020} consists in classifying spoken digits from $0$ to $9$ while involving various levels of non-linearity in the pre-treatment of the data. Indeed, the audio signals can be pre-processed using acoustic filters with different degrees of non-linearity in order to ease their further classification by the neural network. These acoustic filters allow to non-linearly extract acoustic features from the input signal by decomposing it into a given set of frequency channels. %
Among the non-linear filters used, the Mel-frequency cepstral coefficients (MFCC) and Lyon's cochlear model (cochleagram) are based on mimicking the filtering that occurs biologically. The third non-linear filter, Spectro HP, is a custom non-linear transformation of the complex decomposition of the data.
The main result of this study is the fact that when the level of non-linearity of the acoustic filter applied on the input data increases, the contribution of the STVO-based neural network in the processing of the data decreases. This reinforces the idea that the effective role of the neural network is to process the data non-linearly, and that standalone non-linear acoustic filters can already achieve high levels of recognition accuracy without the help of a neural network~\cite{abreu_araujo_riou_torrejon_tsunegi_querlioz_yakushiji_fukushima_kubota_yuasa_stiles_2020}. %
It was also observed that in some cases, the simulated results (with the phenomenological model based on the non-linear magnetic oscillator theory) were surprisingly lower than the experimental ones, as the simulations do not suffer from experimental conditions that may lead to some inaccuracies. This motivates the development of the more accurate physical models presented in this work.

\begin{figure*}
\centering%
\includegraphics[width=.65\textwidth]{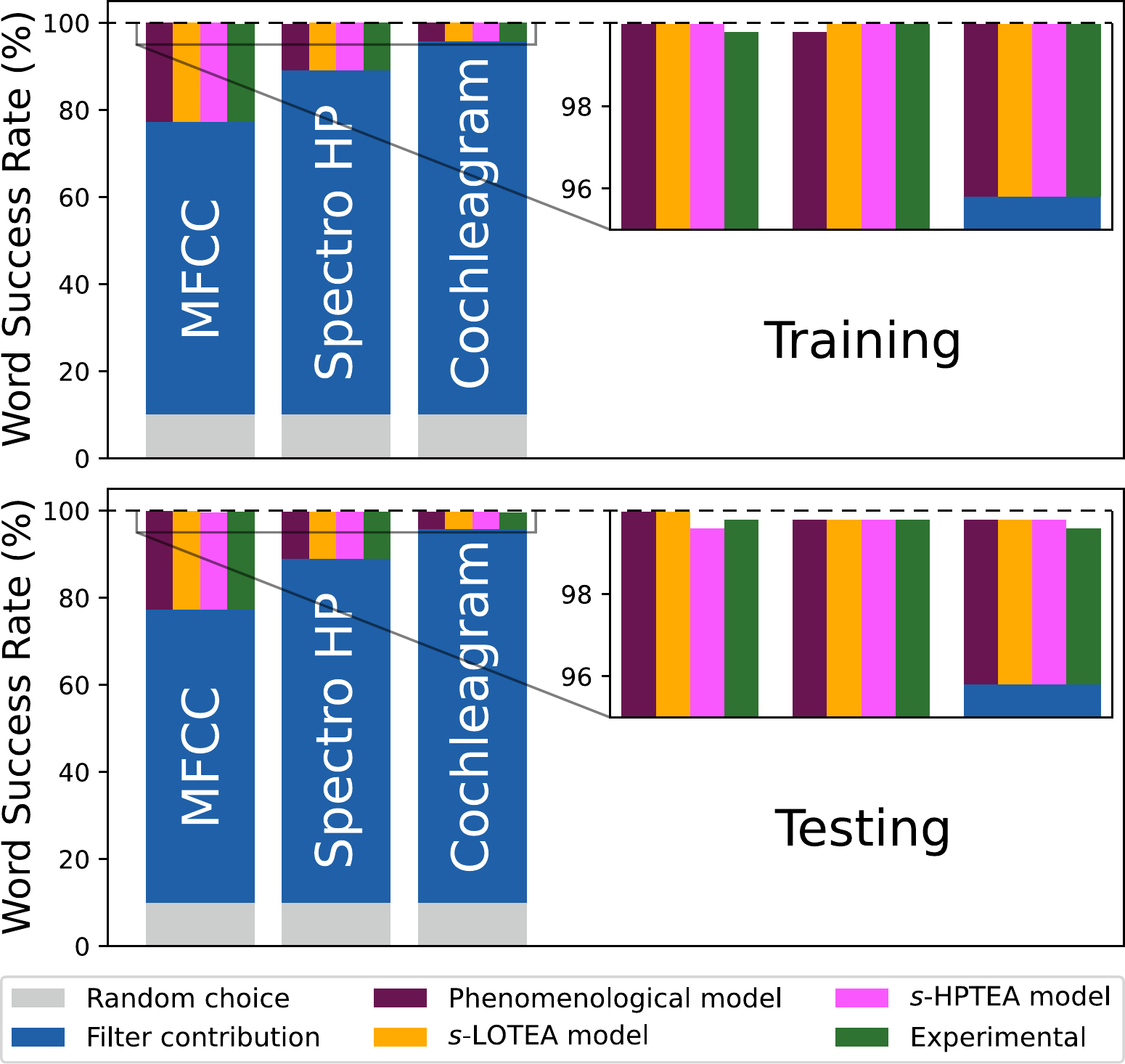}
\caption{Comparison between the different contributions to the accuracy (or Word-Success-Rate, WSR), obtained with the WTA approach during the training (top) and testing (bottom) phases, with different acoustic filters. The green columns are from \myRef~\cite{abreu_araujo_riou_torrejon_tsunegi_querlioz_yakushiji_fukushima_kubota_yuasa_stiles_2020}. The purple, orange and pink columns represent respectively the accuracy obtained with the phenomenological model, the $s$-LOTEA model, and the $s$-HPTEA model.}\label{fig:speech_2020}
\end{figure*}

The results obtained for the training and testing phases are displayed in \fig\ref{fig:speech_2020}. In general, a very good agreement between the experimental results and the the results obtained with our two analytical models is observed. This study hence allows us to conclude on the consistency of our models with the experimental reality at a higher level than just the value of the reduced position of the vortex core $s(t)$. 

It was noticed that the accuracy was systematically higher during the classification of the averaged $V_\text{AVG16}$ signals than for the raw $V_\text{AVG1}$ signals, and the RMSE systematically lower. The only cases where the performance are not improving is when the accuracy is already equal to $100\%$ with the $V_\text{AVG1}$ signals, or when the RMSE has reached a plateau such as the one showed at high SNRs in \fig\ref{fig:snr}. These observations validate the earlier results obtained from the parametric sweep of the SNR, \textit{i.e.} the quality of the recognition improves significantly at higher SNRs.

\begin{table}
    \caption{Estimated SNR for the experimental measurements using different non-linear acoustic filters for speech recognition (data from \myRef~\cite{abreu_araujo_riou_torrejon_tsunegi_querlioz_yakushiji_fukushima_kubota_yuasa_stiles_2020}). }
    \centering%
    \begin{ruledtabular}
    \begin{tabular}{lr}
         \textbf{Non-linear filter}& \textbf{Estimation of the SNR ($\widehat{SNR}$)}\\
         \hline
         MFCC & $18.76$ dB\\
         Cochleagram & $15.59$ dB\\
         Spectro HP & $19.78$ dB\\
    \end{tabular}
    \end{ruledtabular}\label{tab:snrs}
\end{table}

\subsection{Image recognition}
Although it is accepted that reservoir computing is usually not suitable for solving non-temporal problems, our system has been successfully tested on the MNIST written digits dataset~\cite{deng2012mnist}. The dimension of the images, which is initially $28\times 28$ pixels, was first reduced down to $44$ values corresponding to the most statistically significant components extracted from the training data using principal components analysis (PCA)~\cite{wold1987principal, abdi2013computational}. These components allow to explain more than $80\%$ of the variance in the training images and are then used to extract useful features from the testing data.
Such a proper feature extraction step is required to make the data more explicit but also to improve the energy efficiency of the training phase thanks to the dimensionality reduction. However, the principal components only have to be retrieved once, and this step can be performed offline beforehand. Moreover, the PCA technique is linear unlike the acoustic filters presented in the framework of speech recognition. Hence, the contribution of the STVO-based reservoir stays relevant as it consists of the non-linear treatment of the data.

The weights of a reservoir composed of $5000$ virtual neurons were trained on the $60000$ feature-extracted images of the training dataset. The inference phase was then performed on the $10000$ images of the testing dataset.
As in the previous tasks, the virtual neurons were implemented using a single STVO under the time multiplexing scheme. The output of the STVO was computed using our $s$-LOTEA model presented in \eq(\ref{eq:lotea}). %

In the vast majority of the cases, the simulations gave rise to an accuracy above $98\%$, with some simulations achieving an accuracy of $98.8\%$. Those results fall well into the state-of-the-art performance for this task~\cite{mnist_benchmark}, while staying lower than what is achievable using a standard architecture such as a convolutional neural network (CNN)~\cite{an2020ensemble}. One has however to keep in mind that the system we propose is ready to be implemented in hardware, which is not possible at the moment for more conventional deep architectures. 
Moreover, the training process is relatively quick~\footnote{The training of a $5000$-nodes neural reservoir on the whole training set took in average $47.3$ seconds on a Dell Precision 5820 tower equipped with an Intel Core i9-10980XE processor ($18$ cores and up to $36$ threads), and $64.2$ seconds on a MacBook Pro equipped with the M2 chip (12 cores CPU).} and is more straightforward than for conventional software implementations. Indeed the key step of the training process only consists in the Moore-Penrose pseudo-inversion of a matrix, a problem that is already well described in the literature~\cite{courrieu2008fast, barata2012moore}. Note that the average performances of our model are above some of the recent hardware implementations of the MNIST resolution~\cite{du2017reservoir,tu2023neural}.

\begin{figure}
    \centering
    \includegraphics[width=.47\textwidth]{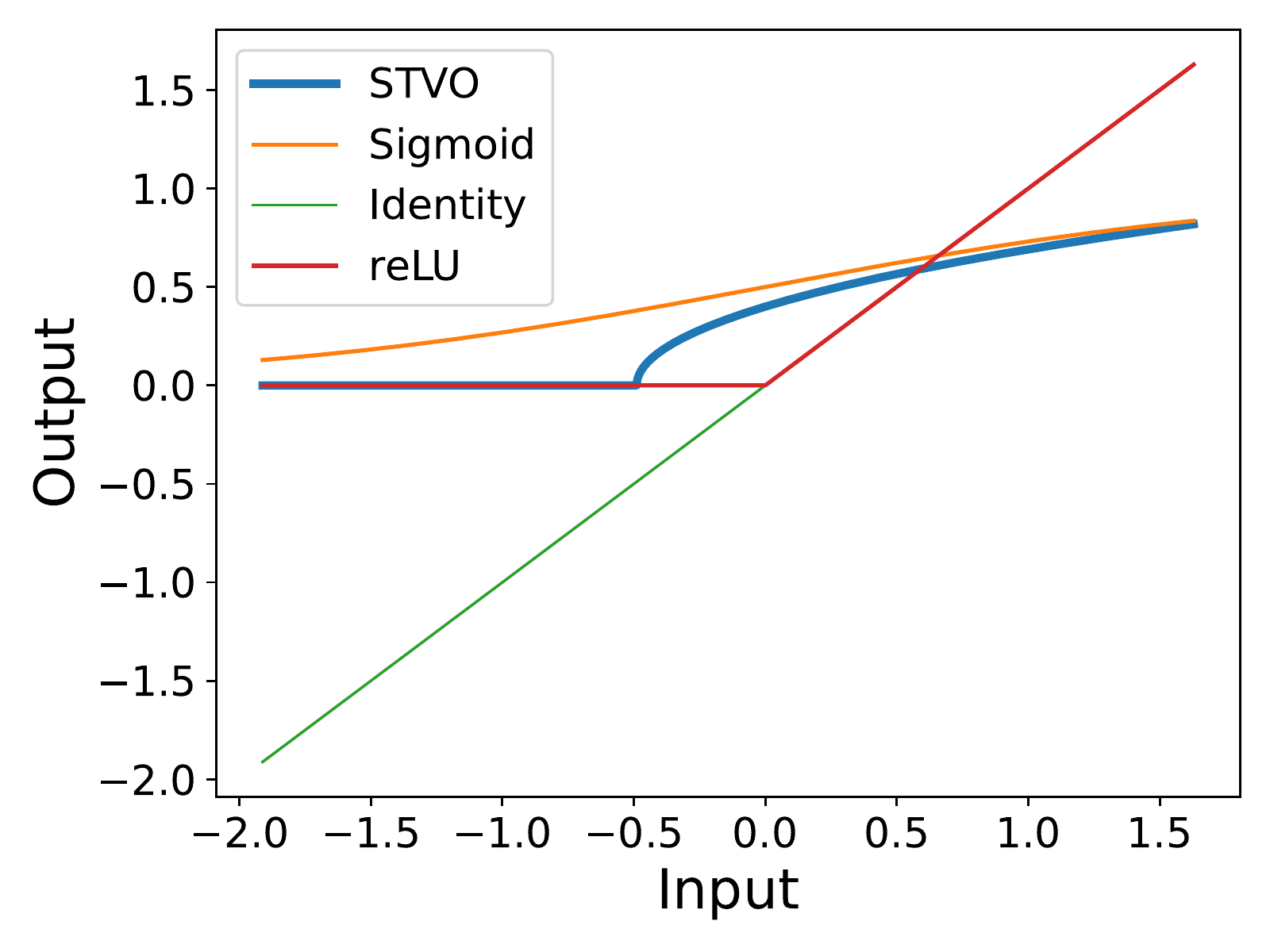}
    \caption{Activation functions benchmarked during the MNIST resolution.}
    \label{fig:act_funs}
\end{figure}

\begin{figure}
    \centering
    \includegraphics[width=.47\textwidth]{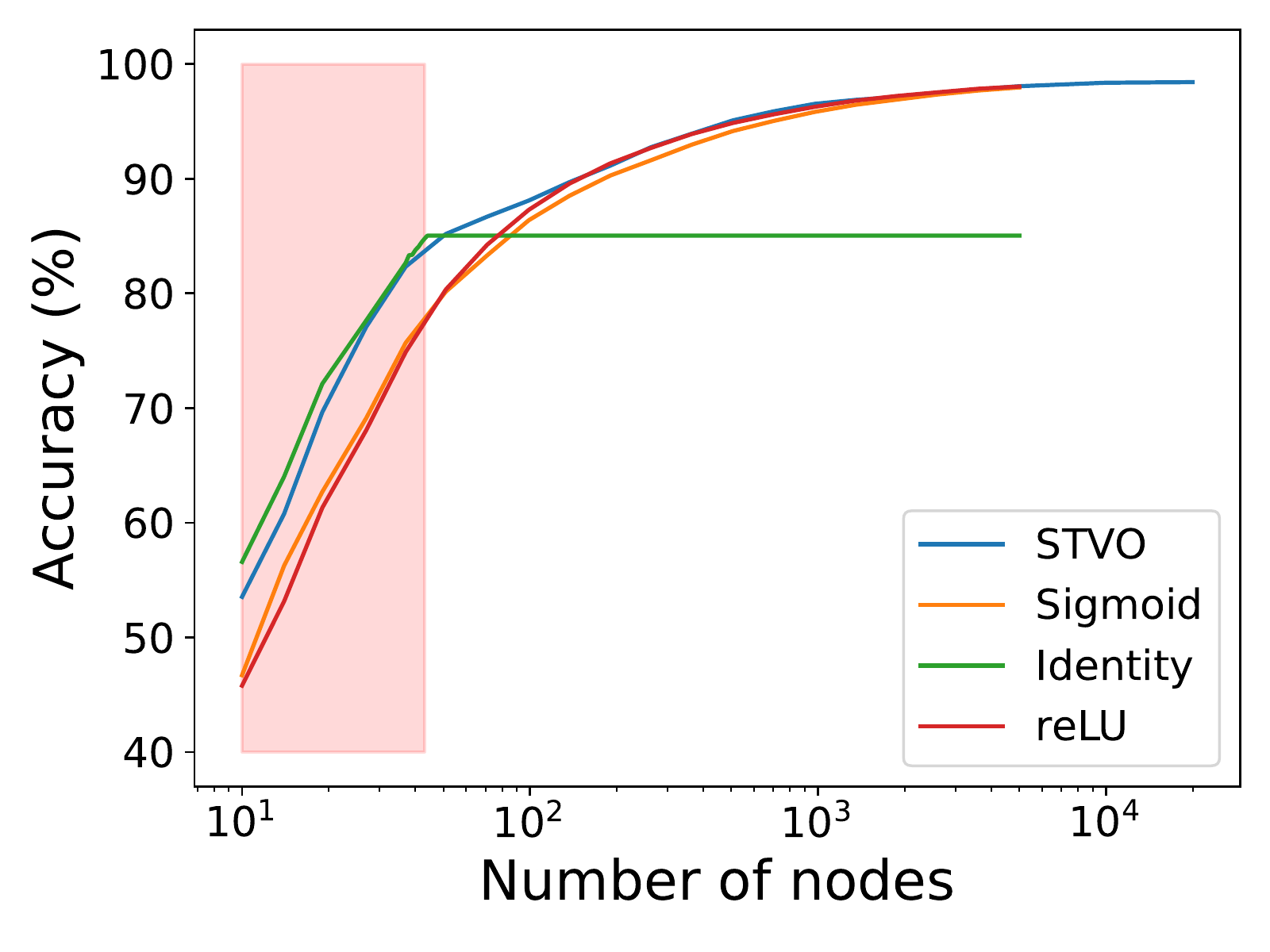}
    \caption{Accuracy of the recognition on the MNIST dataset depending on the number of virtual nodes in the reservoir for different activation functions. The red shaded area represents the underfitted configurations.}
    \label{fig:func_benchmark}
\end{figure}

The versatility of our simulations allowed to compare the vortex core reduced position $s_\infty(J)$ presented in \fig\ref{fig:map} and \fig\ref{fig:map2} with more conventional activation functions such as the reLU and the sigmoid (see \fig\ref{fig:act_funs} and \fig\ref{fig:func_benchmark}). When the number of virtual nodes in the network is lower than $44$, the model described by the network is underfitted : the number of free parameters in the network is not sufficient to fit the dimension of the input data, and hence the information is compressed. As the number of nodes increases towards $44$, the model achieves an increasingly better fit of the data. We believe that below $44$ nodes, the non-linearity of the activation functions is not used to treat the data and can even be detrimental for the recognition due to additional data compression (for example, all negatives inputs are mapped to $0$ with the reLU, see \fig\ref{fig:act_funs}). This would explain why the identity function achieves an accuracy higher than the ReLU and the sigmoid below $44$ nodes. The STVO function still reaches a better accuracy than the reLU and the sigmoid in this case (\fig\ref{fig:func_benchmark}).

Once the number of nodes in the network exceeds $44$, the non-linearity of the activation functions becomes decisive. The identity function saturates at $~85\%$, which is the accuracy that one can expect to obtain when using a simple linear regression on the first $44$ PCA components. That clearly shows the importance of non-linearity in the treatment of the data. For the other functions on the other hand, the intrinsic non-linearity allows to reach higher accuracy levels. One can observe that the reached accuracy tends to the same value for the STVO, reLU and sigmoid functions, implying that the shape of the non-linearity is no longer important for a high level of nodes. However, there is a range between $44$ and $150$ nodes where the STVO function performs better than the two other ones. We suspect that the shape of the STVO non-linearity (blue curve in \fig\ref{fig:act_funs}) brings together the characteristics of the reLU function (\textit{i.e.} all inputs are mapped to $0$ under a certain threshold) and some characteristics of the sigmoid function (\textit{i.e.} a non-linear monotonically increasing part).

Furthermore, the STVO function is highly tunable through the parameters $\Delta V$, $D_\text{t}$ and $I_\text{w}$. This brings the perspective of designing activation functions that are specific to the task one wants to perform, using the same device. The investigation of such \textit{ad hoc} activation functions is made possible thanks to the high throughput of the simulations based on our two new analytical models.  

\section{Conclusion}
Two new models based on the Thiele equation approach were developed to simulate the dynamics of STVOs. The combination of mathematical expressions with numerical results obtained with MMS allowed to obtain a physical description of the vortex core dynamics under a given input signal. The physical accuracy of our new framework is the same as MMS\@. Additional mathematical and numerical adjustments were brought to allow the analytical resolution of the models and accelerate the simulations up to $9$ orders of magnitude compared to MMS\@. One of the model ($s$-LOTEA) truncates the transient dynamic factor $\Gamma(s)$ representing the non-linearity of the dynamics to the second order. %
The other model ($s$-HPTEA) allows to take the whole complexity of the dynamics non-linearity into account. These models were used to simulate a STVO, which led to the design of various virtual STVO-based recurrent neural networks (reservoirs) multiplexed through time.

We first successfully classified sine and square periods automatically. The consideration of past inputs in our new models allows to simulate the intrinsic short-term memory of STVOs. This helps to discriminate the different waveforms even when the value of the signals is identical at some points. The speed of the simulations allowed to perform two parametric studies about the performances of the network during this task, with respect to the working current intensity $I_\text{w}$ and the level of noise in the system. %

The first study allowed to observe that the intrinsic STVO behavior allows a better data treatment at higher current intensities, and to determine an optimal working current intensity for the use of an experimental setup for this given task. The second study allowed to drawn unprecedented relations between the accuracy of the classification and the signal-to-noise ratio of the system. These two studies would have been impossible to perform in practice using MMS due to the extensive required simulation time, hence highlighting the value of our analytical models. %

We then performed spoken digits recognition on the TI-$46$ dataset in order to compare the results with the values obtained experimentally in \myRef~\cite{torrejon_riou_araujo_tsunegi_khalsa_querlioz_bortolotti_cros_yakushiji_fukushima_2017, abreu_araujo_riou_torrejon_tsunegi_querlioz_yakushiji_fukushima_kubota_yuasa_stiles_2020}. A very good agreement was obtained between our simulated results and the experimental results, hence showing the consistency of the simulations based on our new models with the physical reality at a higher applicative level. An improvement of the recognition quality with the SNR of the system is also observed experimentally, validating the conclusions made at the end of the second parametric study presented earlier. 

Finally, the generalizability of our solution to non-temporal recognition tasks was demonstrated by reaching state-of-the-art accuracy levels on the MNIST dataset within a relatively short computation time. While staying below the accuracy levels obtained with more conventional software architectures, our proposed solution has the advantage to be ready to be implemented in hardware. Moreover, we showed that the STVO dynamics can be used as an \textit{ad hoc} activation function by tuning the operating parameters of the STVO, and that it performed better than the reLU and the sigmoid functions when the number of free parameters in the network is between $44$ and $150$.

The use of our new analytical models based on the physical description of STVO dynamics is expected to lead to a major development in the simulation of STVO-based neural networks. The huge acceleration factor induced by their analytical resolution will allow to simulate larger and more complex networks, opening the path to novel neuromorphic schemes and STVO-based architectures.

\section{Acknowledgements}
Computational resources have been provided by the Consortium des Équipements de Calcul Intensif (CÉCI), funded by the Fonds de la Recherche Scientifique de Belgique (F.R.S.-FNRS) under Grant No. 2.5020.11 and by the Walloon Region. F.A.A. is a Research Associate and S.d.W. is a FRIA grantee, both of the F.R.S.-FNRS\@. %
The experimental data originates from the measurements by Jacob Torrejon, Mathieu Riou, and F.A.A. while working in Julie Grollier's group at Unité Mixte de Physique CNRS/Thales in France and the samples were produced at AIST (Spintronics Research Center, Tsukuba) in Japan under the responsibility of our colleagues Sumito Tsunegi, Kay Yakushiji, Akio Fukushima, Hitoshi Kubota, and Shinji Yuasa.

\section{Author's contribution}
The study was designed by F.A.A. who created the analytical foundation related to the Thiele equation approach with the assistance of S.d.W. and C.C.. F.A.A. developed the two analytical models and combined them with numerical data obtained by micromagnetic simulations performed by S.d.W.. A.M. designed and performed the benchmark tasks, the parametric studies, and compared the numerical results with the experiment data. A.M. wrote the core of the manuscript and all the other co-authors (F.A.A., S.d.W., C.C., and J.W.) contributed to the text as well as to the analysis of the results.

\section{References}

\bibliography{biblio}

\begin{thebibliography}{38}%
\makeatletter
\providecommand \@ifxundefined [1]{%
 \@ifx{#1\undefined}
}%
\providecommand \@ifnum [1]{%
 \ifnum #1\expandafter \@firstoftwo
 \else \expandafter \@secondoftwo
 \fi
}%
\providecommand \@ifx [1]{%
 \ifx #1\expandafter \@firstoftwo
 \else \expandafter \@secondoftwo
 \fi
}%
\providecommand \natexlab [1]{#1}%
\providecommand \enquote  [1]{``#1''}%
\providecommand \bibnamefont  [1]{#1}%
\providecommand \bibfnamefont [1]{#1}%
\providecommand \citenamefont [1]{#1}%
\providecommand \href@noop [0]{\@secondoftwo}%
\providecommand \href [0]{\begingroup \@sanitize@url \@href}%
\providecommand \@href[1]{\@@startlink{#1}\@@href}%
\providecommand \@@href[1]{\endgroup#1\@@endlink}%
\providecommand \@sanitize@url [0]{\catcode `\\12\catcode `\$12\catcode
  `\&12\catcode `\#12\catcode `\^12\catcode `\_12\catcode `\%12\relax}%
\providecommand \@@startlink[1]{}%
\providecommand \@@endlink[0]{}%
\providecommand \url  [0]{\begingroup\@sanitize@url \@url }%
\providecommand \@url [1]{\endgroup\@href {#1}{\urlprefix }}%
\providecommand \urlprefix  [0]{URL }%
\providecommand \Eprint [0]{\href }%
\providecommand \doibase [0]{http://dx.doi.org/}%
\providecommand \selectlanguage [0]{\@gobble}%
\providecommand \bibinfo  [0]{\@secondoftwo}%
\providecommand \bibfield  [0]{\@secondoftwo}%
\providecommand \translation [1]{[#1]}%
\providecommand \BibitemOpen [0]{}%
\providecommand \bibitemStop [0]{}%
\providecommand \bibitemNoStop [0]{.\EOS\space}%
\providecommand \EOS [0]{\spacefactor3000\relax}%
\providecommand \BibitemShut  [1]{\csname bibitem#1\endcsname}%
\let\auto@bib@innerbib\@empty
\bibitem [{\citenamefont {Indiveri}\ and\ \citenamefont
  {Horiuchi}(2011)}]{indiveri2011frontiers}%
  \BibitemOpen
  \bibfield  {author} {\bibinfo {author} {\bibfnamefont {G.}~\bibnamefont
  {Indiveri}}\ and\ \bibinfo {author} {\bibfnamefont {T.~K.}\ \bibnamefont
  {Horiuchi}},\ }\href {\doibase 10.3389/fnins.2011.00118} {\bibfield
  {journal} {\bibinfo  {journal} {Frontiers in neuroscience}\ }\textbf
  {\bibinfo {volume} {5}},\ \bibinfo {pages} {118} (\bibinfo {year}
  {2011})}\BibitemShut {NoStop}%
\bibitem [{\citenamefont {Paquot}\ \emph {et~al.}(2010)\citenamefont {Paquot},
  \citenamefont {Dambre}, \citenamefont {Schrauwen}, \citenamefont
  {Haelterman},\ and\ \citenamefont {Massar}}]{paquot2010reservoir}%
  \BibitemOpen
  \bibfield  {author} {\bibinfo {author} {\bibfnamefont {Y.}~\bibnamefont
  {Paquot}}, \bibinfo {author} {\bibfnamefont {J.}~\bibnamefont {Dambre}},
  \bibinfo {author} {\bibfnamefont {B.}~\bibnamefont {Schrauwen}}, \bibinfo
  {author} {\bibfnamefont {M.}~\bibnamefont {Haelterman}}, \ and\ \bibinfo
  {author} {\bibfnamefont {S.}~\bibnamefont {Massar}},\ }in\ \href@noop {}
  {\emph {\bibinfo {booktitle} {Nonlinear optics and applications IV}}},\ Vol.\
  \bibinfo {volume} {7728}\ (\bibinfo {organization} {SPIE},\ \bibinfo {year}
  {2010})\ pp.\ \bibinfo {pages} {58--69}\BibitemShut {NoStop}%
\bibitem [{\citenamefont {Paquot}\ \emph {et~al.}(2012)\citenamefont {Paquot},
  \citenamefont {Duport}, \citenamefont {Smerieri}, \citenamefont {Dambre},
  \citenamefont {Schrauwen}, \citenamefont {Haelterman},\ and\ \citenamefont
  {Massar}}]{paquot2012optoelectronic}%
  \BibitemOpen
  \bibfield  {author} {\bibinfo {author} {\bibfnamefont {Y.}~\bibnamefont
  {Paquot}}, \bibinfo {author} {\bibfnamefont {F.}~\bibnamefont {Duport}},
  \bibinfo {author} {\bibfnamefont {A.}~\bibnamefont {Smerieri}}, \bibinfo
  {author} {\bibfnamefont {J.}~\bibnamefont {Dambre}}, \bibinfo {author}
  {\bibfnamefont {B.}~\bibnamefont {Schrauwen}}, \bibinfo {author}
  {\bibfnamefont {M.}~\bibnamefont {Haelterman}}, \ and\ \bibinfo {author}
  {\bibfnamefont {S.}~\bibnamefont {Massar}},\ }\href@noop {} {\bibfield
  {journal} {\bibinfo  {journal} {Scientific Reports}\ }\textbf {\bibinfo
  {volume} {2}},\ \bibinfo {pages} {1} (\bibinfo {year} {2012})}\BibitemShut
  {NoStop}%
\bibitem [{\citenamefont {Larger}\ \emph {et~al.}(2012)\citenamefont {Larger},
  \citenamefont {Soriano}, \citenamefont {Brunner}, \citenamefont {Appeltant},
  \citenamefont {Guti{\'e}rrez}, \citenamefont {Pesquera}, \citenamefont
  {Mirasso},\ and\ \citenamefont {Fischer}}]{larger2012photonic}%
  \BibitemOpen
  \bibfield  {author} {\bibinfo {author} {\bibfnamefont {L.}~\bibnamefont
  {Larger}}, \bibinfo {author} {\bibfnamefont {M.~C.}\ \bibnamefont {Soriano}},
  \bibinfo {author} {\bibfnamefont {D.}~\bibnamefont {Brunner}}, \bibinfo
  {author} {\bibfnamefont {L.}~\bibnamefont {Appeltant}}, \bibinfo {author}
  {\bibfnamefont {J.~M.}\ \bibnamefont {Guti{\'e}rrez}}, \bibinfo {author}
  {\bibfnamefont {L.}~\bibnamefont {Pesquera}}, \bibinfo {author}
  {\bibfnamefont {C.~R.}\ \bibnamefont {Mirasso}}, \ and\ \bibinfo {author}
  {\bibfnamefont {I.}~\bibnamefont {Fischer}},\ }\href@noop {} {\bibfield
  {journal} {\bibinfo  {journal} {Optics Express}\ }\textbf {\bibinfo {volume}
  {20}},\ \bibinfo {pages} {3241} (\bibinfo {year} {2012})}\BibitemShut
  {NoStop}%
\bibitem [{\citenamefont {Larger}\ \emph {et~al.}(2017)\citenamefont {Larger},
  \citenamefont {Bayl{\'o}n-Fuentes}, \citenamefont {Martinenghi},
  \citenamefont {Udaltsov}, \citenamefont {Chembo},\ and\ \citenamefont
  {Jacquot}}]{larger2017high}%
  \BibitemOpen
  \bibfield  {author} {\bibinfo {author} {\bibfnamefont {L.}~\bibnamefont
  {Larger}}, \bibinfo {author} {\bibfnamefont {A.}~\bibnamefont
  {Bayl{\'o}n-Fuentes}}, \bibinfo {author} {\bibfnamefont {R.}~\bibnamefont
  {Martinenghi}}, \bibinfo {author} {\bibfnamefont {V.~S.}\ \bibnamefont
  {Udaltsov}}, \bibinfo {author} {\bibfnamefont {Y.~K.}\ \bibnamefont
  {Chembo}}, \ and\ \bibinfo {author} {\bibfnamefont {M.}~\bibnamefont
  {Jacquot}},\ }\href@noop {} {\bibfield  {journal} {\bibinfo  {journal}
  {Physical Review X}\ }\textbf {\bibinfo {volume} {7}},\ \bibinfo {pages}
  {011015} (\bibinfo {year} {2017})}\BibitemShut {NoStop}%
\bibitem [{\citenamefont {Shastri}\ \emph {et~al.}(2021)\citenamefont
  {Shastri}, \citenamefont {Tait}, \citenamefont {de~Lima}, \citenamefont
  {Pernice}, \citenamefont {Bhaskaran}, \citenamefont {Wright},\ and\
  \citenamefont {Prucnal}}]{shastri2021photonics}%
  \BibitemOpen
  \bibfield  {author} {\bibinfo {author} {\bibfnamefont {B.~J.}\ \bibnamefont
  {Shastri}}, \bibinfo {author} {\bibfnamefont {A.~N.}\ \bibnamefont {Tait}},
  \bibinfo {author} {\bibfnamefont {T.~F.}\ \bibnamefont {de~Lima}}, \bibinfo
  {author} {\bibfnamefont {W.~H.}\ \bibnamefont {Pernice}}, \bibinfo {author}
  {\bibfnamefont {H.}~\bibnamefont {Bhaskaran}}, \bibinfo {author}
  {\bibfnamefont {C.~D.}\ \bibnamefont {Wright}}, \ and\ \bibinfo {author}
  {\bibfnamefont {P.~R.}\ \bibnamefont {Prucnal}},\ }\href {\doibase
  10.1038/s41566-020-00754-y} {\bibfield  {journal} {\bibinfo  {journal}
  {Nature Photonics}\ }\textbf {\bibinfo {volume} {15}},\ \bibinfo {pages}
  {102} (\bibinfo {year} {2021})}\BibitemShut {NoStop}%
\bibitem [{\citenamefont {Jo}\ \emph {et~al.}(2010)\citenamefont {Jo},
  \citenamefont {Chang}, \citenamefont {Ebong}, \citenamefont {Bhadviya},
  \citenamefont {Mazumder},\ and\ \citenamefont {Lu}}]{jo2010nanoscale}%
  \BibitemOpen
  \bibfield  {author} {\bibinfo {author} {\bibfnamefont {S.~H.}\ \bibnamefont
  {Jo}}, \bibinfo {author} {\bibfnamefont {T.}~\bibnamefont {Chang}}, \bibinfo
  {author} {\bibfnamefont {I.}~\bibnamefont {Ebong}}, \bibinfo {author}
  {\bibfnamefont {B.~B.}\ \bibnamefont {Bhadviya}}, \bibinfo {author}
  {\bibfnamefont {P.}~\bibnamefont {Mazumder}}, \ and\ \bibinfo {author}
  {\bibfnamefont {W.}~\bibnamefont {Lu}},\ }\href {\doibase 10.1021/nl904092h}
  {\bibfield  {journal} {\bibinfo  {journal} {Nano letters}\ }\textbf {\bibinfo
  {volume} {10}},\ \bibinfo {pages} {1297} (\bibinfo {year}
  {2010})}\BibitemShut {NoStop}%
\bibitem [{\citenamefont {Du}\ \emph {et~al.}(2017)\citenamefont {Du},
  \citenamefont {Cai}, \citenamefont {Zidan}, \citenamefont {Ma}, \citenamefont
  {Lee},\ and\ \citenamefont {Lu}}]{du2017reservoir}%
  \BibitemOpen
  \bibfield  {author} {\bibinfo {author} {\bibfnamefont {C.}~\bibnamefont
  {Du}}, \bibinfo {author} {\bibfnamefont {F.}~\bibnamefont {Cai}}, \bibinfo
  {author} {\bibfnamefont {M.~A.}\ \bibnamefont {Zidan}}, \bibinfo {author}
  {\bibfnamefont {W.}~\bibnamefont {Ma}}, \bibinfo {author} {\bibfnamefont
  {S.~H.}\ \bibnamefont {Lee}}, \ and\ \bibinfo {author} {\bibfnamefont
  {W.~D.}\ \bibnamefont {Lu}},\ }\href@noop {} {\bibfield  {journal} {\bibinfo
  {journal} {Nature communications}\ }\textbf {\bibinfo {volume} {8}},\
  \bibinfo {pages} {2204} (\bibinfo {year} {2017})}\BibitemShut {NoStop}%
\bibitem [{\citenamefont {Grollier}\ \emph {et~al.}(2020)\citenamefont
  {Grollier}, \citenamefont {Querlioz}, \citenamefont {Camsari}, \citenamefont
  {Everschor-Sitte}, \citenamefont {Fukami},\ and\ \citenamefont
  {Stiles}}]{grollier2020neuromorphic}%
  \BibitemOpen
  \bibfield  {author} {\bibinfo {author} {\bibfnamefont {J.}~\bibnamefont
  {Grollier}}, \bibinfo {author} {\bibfnamefont {D.}~\bibnamefont {Querlioz}},
  \bibinfo {author} {\bibfnamefont {K.}~\bibnamefont {Camsari}}, \bibinfo
  {author} {\bibfnamefont {K.}~\bibnamefont {Everschor-Sitte}}, \bibinfo
  {author} {\bibfnamefont {S.}~\bibnamefont {Fukami}}, \ and\ \bibinfo {author}
  {\bibfnamefont {M.~D.}\ \bibnamefont {Stiles}},\ }\href {\doibase
  10.1038/s41928-019-0360-9} {\bibfield  {journal} {\bibinfo  {journal} {Nature
  electronics}\ }\textbf {\bibinfo {volume} {3}},\ \bibinfo {pages} {360}
  (\bibinfo {year} {2020})}\BibitemShut {NoStop}%
\bibitem [{\citenamefont {Torrejon}\ \emph {et~al.}(2017)\citenamefont
  {Torrejon}, \citenamefont {Riou}, \citenamefont {{\relax Abreu Araujo}},
  \citenamefont {Tsunegi}, \citenamefont {Khalsa}, \citenamefont {Querlioz},
  \citenamefont {Bortolotti}, \citenamefont {Cros}, \citenamefont {Yakushiji},
  \citenamefont {Fukushima} \emph
  {et~al.}}]{torrejon_riou_araujo_tsunegi_khalsa_querlioz_bortolotti_cros_yakushiji_fukushima_2017}%
  \BibitemOpen
  \bibfield  {author} {\bibinfo {author} {\bibfnamefont {J.}~\bibnamefont
  {Torrejon}}, \bibinfo {author} {\bibfnamefont {M.}~\bibnamefont {Riou}},
  \bibinfo {author} {\bibfnamefont {F.}~\bibnamefont {{\relax Abreu Araujo}}},
  \bibinfo {author} {\bibfnamefont {S.}~\bibnamefont {Tsunegi}}, \bibinfo
  {author} {\bibfnamefont {G.}~\bibnamefont {Khalsa}}, \bibinfo {author}
  {\bibfnamefont {D.}~\bibnamefont {Querlioz}}, \bibinfo {author}
  {\bibfnamefont {P.}~\bibnamefont {Bortolotti}}, \bibinfo {author}
  {\bibfnamefont {V.}~\bibnamefont {Cros}}, \bibinfo {author} {\bibfnamefont
  {K.}~\bibnamefont {Yakushiji}}, \bibinfo {author} {\bibfnamefont
  {A.}~\bibnamefont {Fukushima}},  \emph {et~al.},\ }\href {\doibase
  10.1038/nature23011} {\bibfield  {journal} {\bibinfo  {journal} {Nature}\
  }\textbf {\bibinfo {volume} {547}},\ \bibinfo {pages} {428–431} (\bibinfo
  {year} {2017})}\BibitemShut {NoStop}%
\bibitem [{\citenamefont {Riou}\ \emph {et~al.}(2019)\citenamefont {Riou},
  \citenamefont {Torrejon}, \citenamefont {Garitaine}, \citenamefont {{\relax
  Abreu Araujo}}, \citenamefont {Bortolotti}, \citenamefont {Cros},
  \citenamefont {Tsunegi}, \citenamefont {Yakushiji}, \citenamefont
  {Fukushima}, \citenamefont {Kubota} \emph
  {et~al.}}]{riou_torrejon_garitaine_abreu_2019}%
  \BibitemOpen
  \bibfield  {author} {\bibinfo {author} {\bibfnamefont {M.}~\bibnamefont
  {Riou}}, \bibinfo {author} {\bibfnamefont {J.}~\bibnamefont {Torrejon}},
  \bibinfo {author} {\bibfnamefont {B.}~\bibnamefont {Garitaine}}, \bibinfo
  {author} {\bibfnamefont {F.}~\bibnamefont {{\relax Abreu Araujo}}}, \bibinfo
  {author} {\bibfnamefont {P.}~\bibnamefont {Bortolotti}}, \bibinfo {author}
  {\bibfnamefont {V.}~\bibnamefont {Cros}}, \bibinfo {author} {\bibfnamefont
  {S.}~\bibnamefont {Tsunegi}}, \bibinfo {author} {\bibfnamefont
  {K.}~\bibnamefont {Yakushiji}}, \bibinfo {author} {\bibfnamefont
  {A.}~\bibnamefont {Fukushima}}, \bibinfo {author} {\bibfnamefont
  {H.}~\bibnamefont {Kubota}},  \emph {et~al.},\ }\href {\doibase
  10.1103/physrevapplied.12.024049} {\bibfield  {journal} {\bibinfo  {journal}
  {Physical Review Applied}\ }\textbf {\bibinfo {volume} {12}} (\bibinfo {year}
  {2019}),\ 10.1103/physrevapplied.12.024049}\BibitemShut {NoStop}%
\bibitem [{\citenamefont {Marković}\ \emph {et~al.}(2019)\citenamefont
  {Marković}, \citenamefont {Leroux}, \citenamefont {Riou}, \citenamefont
  {{\relax Abreu Araujo}}, \citenamefont {Torrejon}, \citenamefont {Querlioz},
  \citenamefont {Fukushima}, \citenamefont {Yuasa}, \citenamefont {Trastoy},
  \citenamefont {Bortolotti} \emph
  {et~al.}}]{markovic_leroux_riou_abreu_araujo_torrejon_querlioz_fukushima_yuasa_trastoy_bortolotti_2019}%
  \BibitemOpen
  \bibfield  {author} {\bibinfo {author} {\bibfnamefont {D.}~\bibnamefont
  {Marković}}, \bibinfo {author} {\bibfnamefont {N.}~\bibnamefont {Leroux}},
  \bibinfo {author} {\bibfnamefont {M.}~\bibnamefont {Riou}}, \bibinfo {author}
  {\bibfnamefont {F.}~\bibnamefont {{\relax Abreu Araujo}}}, \bibinfo {author}
  {\bibfnamefont {J.}~\bibnamefont {Torrejon}}, \bibinfo {author}
  {\bibfnamefont {D.}~\bibnamefont {Querlioz}}, \bibinfo {author}
  {\bibfnamefont {A.}~\bibnamefont {Fukushima}}, \bibinfo {author}
  {\bibfnamefont {S.}~\bibnamefont {Yuasa}}, \bibinfo {author} {\bibfnamefont
  {J.}~\bibnamefont {Trastoy}}, \bibinfo {author} {\bibfnamefont
  {P.}~\bibnamefont {Bortolotti}},  \emph {et~al.},\ }\href {\doibase
  10.1063/1.5079305} {\bibfield  {journal} {\bibinfo  {journal} {Applied
  Physics Letters}\ }\textbf {\bibinfo {volume} {114}},\ \bibinfo {pages}
  {012409} (\bibinfo {year} {2019})}\BibitemShut {NoStop}%
\bibitem [{\citenamefont {Yogendra}\ \emph {et~al.}(2015)\citenamefont
  {Yogendra}, \citenamefont {Fan},\ and\ \citenamefont
  {Roy}}]{yogendra2015lowpower}%
  \BibitemOpen
  \bibfield  {author} {\bibinfo {author} {\bibfnamefont {K.}~\bibnamefont
  {Yogendra}}, \bibinfo {author} {\bibfnamefont {D.}~\bibnamefont {Fan}}, \
  and\ \bibinfo {author} {\bibfnamefont {K.}~\bibnamefont {Roy}},\ }\href
  {\doibase 10.1109/tmag.2015.2443042} {\bibfield  {journal} {\bibinfo
  {journal} {IEEE Transactions on Magnetics}\ }\textbf {\bibinfo {volume}
  {51}},\ \bibinfo {pages} {1–9} (\bibinfo {year} {2015})}\BibitemShut
  {NoStop}%
\bibitem [{\citenamefont {Riou}\ \emph {et~al.}(2017)\citenamefont {Riou},
  \citenamefont {Araujo}, \citenamefont {Torrejon}, \citenamefont {Tsunegi},
  \citenamefont {Khalsa}, \citenamefont {Querlioz}, \citenamefont {Bortolotti},
  \citenamefont {Cros}, \citenamefont {Yakushiji}, \citenamefont {Fukushima}
  \emph {et~al.}}]{riou2017neuromorphic}%
  \BibitemOpen
  \bibfield  {author} {\bibinfo {author} {\bibfnamefont {M.}~\bibnamefont
  {Riou}}, \bibinfo {author} {\bibfnamefont {F.~A.}\ \bibnamefont {Araujo}},
  \bibinfo {author} {\bibfnamefont {J.}~\bibnamefont {Torrejon}}, \bibinfo
  {author} {\bibfnamefont {S.}~\bibnamefont {Tsunegi}}, \bibinfo {author}
  {\bibfnamefont {G.}~\bibnamefont {Khalsa}}, \bibinfo {author} {\bibfnamefont
  {D.}~\bibnamefont {Querlioz}}, \bibinfo {author} {\bibfnamefont
  {P.}~\bibnamefont {Bortolotti}}, \bibinfo {author} {\bibfnamefont
  {V.}~\bibnamefont {Cros}}, \bibinfo {author} {\bibfnamefont {K.}~\bibnamefont
  {Yakushiji}}, \bibinfo {author} {\bibfnamefont {A.}~\bibnamefont
  {Fukushima}},  \emph {et~al.},\ }in\ \href@noop {} {\emph {\bibinfo
  {booktitle} {2017 IEEE International Electron Devices Meeting (IEDM)}}}\
  (\bibinfo {organization} {IEEE},\ \bibinfo {year} {2017})\ pp.\ \bibinfo
  {pages} {36--3}\BibitemShut {NoStop}%
\bibitem [{\citenamefont {Riou}\ \emph {et~al.}(2021)\citenamefont {Riou},
  \citenamefont {Torrejon}, \citenamefont {Abreu~Araujo}, \citenamefont
  {Tsunegi}, \citenamefont {Khalsa}, \citenamefont {Querlioz}, \citenamefont
  {Bortolotti}, \citenamefont {Leroux}, \citenamefont {Markovi{\'c}},
  \citenamefont {Cros} \emph {et~al.}}]{riou2021reservoir}%
  \BibitemOpen
  \bibfield  {author} {\bibinfo {author} {\bibfnamefont {M.}~\bibnamefont
  {Riou}}, \bibinfo {author} {\bibfnamefont {J.}~\bibnamefont {Torrejon}},
  \bibinfo {author} {\bibfnamefont {F.}~\bibnamefont {Abreu~Araujo}}, \bibinfo
  {author} {\bibfnamefont {S.}~\bibnamefont {Tsunegi}}, \bibinfo {author}
  {\bibfnamefont {G.}~\bibnamefont {Khalsa}}, \bibinfo {author} {\bibfnamefont
  {D.}~\bibnamefont {Querlioz}}, \bibinfo {author} {\bibfnamefont
  {P.}~\bibnamefont {Bortolotti}}, \bibinfo {author} {\bibfnamefont
  {N.}~\bibnamefont {Leroux}}, \bibinfo {author} {\bibfnamefont
  {D.}~\bibnamefont {Markovi{\'c}}}, \bibinfo {author} {\bibfnamefont
  {V.}~\bibnamefont {Cros}},  \emph {et~al.},\ }in\ \href@noop {} {\emph
  {\bibinfo {booktitle} {Reservoir Computing}}}\ (\bibinfo  {publisher}
  {Springer},\ \bibinfo {year} {2021})\ pp.\ \bibinfo {pages}
  {307--329}\BibitemShut {NoStop}%
\bibitem [{\citenamefont {Vansteenkiste}\ \emph {et~al.}(2014)\citenamefont
  {Vansteenkiste}, \citenamefont {Jonathan}, \citenamefont {Dvornik},
  \citenamefont {Helsen}, \citenamefont {Garcia-Sanchez},\ and\ \citenamefont
  {Van~Waeyenberge}}]{mumax3}%
  \BibitemOpen
  \bibfield  {author} {\bibinfo {author} {\bibfnamefont {A.}~\bibnamefont
  {Vansteenkiste}}, \bibinfo {author} {\bibfnamefont {L.}~\bibnamefont
  {Jonathan}}, \bibinfo {author} {\bibfnamefont {M.}~\bibnamefont {Dvornik}},
  \bibinfo {author} {\bibfnamefont {M.}~\bibnamefont {Helsen}}, \bibinfo
  {author} {\bibfnamefont {F.}~\bibnamefont {Garcia-Sanchez}}, \ and\ \bibinfo
  {author} {\bibfnamefont {B.}~\bibnamefont {Van~Waeyenberge}},\ }\href
  {\doibase 10.1063/1.4899186.} {\bibfield  {journal} {\bibinfo  {journal} {AIP
  Advances}\ }\textbf {\bibinfo {volume} {4}},\ \bibinfo {pages} {107}
  (\bibinfo {year} {2014})}\BibitemShut {NoStop}%
\bibitem [{\citenamefont {Abreu~Araujo}\ \emph
  {et~al.}(2022{\natexlab{a}})\citenamefont {Abreu~Araujo}, \citenamefont
  {Chopin},\ and\ \citenamefont
  {de~Wergifosse}}]{abreu_araujo_chopin_de_wergifosse_2022}%
  \BibitemOpen
  \bibfield  {author} {\bibinfo {author} {\bibfnamefont {F.}~\bibnamefont
  {Abreu~Araujo}}, \bibinfo {author} {\bibfnamefont {C.}~\bibnamefont
  {Chopin}}, \ and\ \bibinfo {author} {\bibfnamefont {S.}~\bibnamefont
  {de~Wergifosse}},\ }\href {\doibase 10.48550/arXiv.2206.13596} {\bibfield
  {journal} {\bibinfo  {journal} {arXiv:2206.13596}\ } (\bibinfo {year}
  {2022}{\natexlab{a}}),\ 10.48550/arXiv.2206.13596}\BibitemShut {NoStop}%
\bibitem [{\citenamefont {Slavin}\ and\ \citenamefont
  {Tiberkevich}(2009)}]{slavin2009nonlinear}%
  \BibitemOpen
  \bibfield  {author} {\bibinfo {author} {\bibfnamefont {A.}~\bibnamefont
  {Slavin}}\ and\ \bibinfo {author} {\bibfnamefont {V.}~\bibnamefont
  {Tiberkevich}},\ }\href {\doibase 10.1109/TMAG.2008.2009935} {\bibfield
  {journal} {\bibinfo  {journal} {IEEE Transactions on Magnetics}\ }\textbf
  {\bibinfo {volume} {45}},\ \bibinfo {pages} {1875} (\bibinfo {year}
  {2009})}\BibitemShut {NoStop}%
\bibitem [{\citenamefont {{\relax Abreu Araujo}}\ \emph
  {et~al.}(2020)\citenamefont {{\relax Abreu Araujo}}, \citenamefont {Riou},
  \citenamefont {Torrejon}, \citenamefont {Tsunegi}, \citenamefont {Querlioz},
  \citenamefont {Yakushiji}, \citenamefont {Fukushima}, \citenamefont {Kubota},
  \citenamefont {Yuasa}, \citenamefont {Stiles} \emph
  {et~al.}}]{abreu_araujo_riou_torrejon_tsunegi_querlioz_yakushiji_fukushima_kubota_yuasa_stiles_2020}%
  \BibitemOpen
  \bibfield  {author} {\bibinfo {author} {\bibfnamefont {F.}~\bibnamefont
  {{\relax Abreu Araujo}}}, \bibinfo {author} {\bibfnamefont {M.}~\bibnamefont
  {Riou}}, \bibinfo {author} {\bibfnamefont {J.}~\bibnamefont {Torrejon}},
  \bibinfo {author} {\bibfnamefont {S.}~\bibnamefont {Tsunegi}}, \bibinfo
  {author} {\bibfnamefont {D.}~\bibnamefont {Querlioz}}, \bibinfo {author}
  {\bibfnamefont {K.}~\bibnamefont {Yakushiji}}, \bibinfo {author}
  {\bibfnamefont {A.}~\bibnamefont {Fukushima}}, \bibinfo {author}
  {\bibfnamefont {H.}~\bibnamefont {Kubota}}, \bibinfo {author} {\bibfnamefont
  {S.}~\bibnamefont {Yuasa}}, \bibinfo {author} {\bibfnamefont {M.~D.}\
  \bibnamefont {Stiles}},  \emph {et~al.},\ }\href {\doibase
  10.1038/s41598-019-56991-x} {\bibfield  {journal} {\bibinfo  {journal}
  {Scientific Reports}\ }\textbf {\bibinfo {volume} {10}} (\bibinfo {year}
  {2020}),\ 10.1038/s41598-019-56991-x}\BibitemShut {NoStop}%
\bibitem [{\citenamefont {Chen}\ \emph {et~al.}(2022)\citenamefont {Chen},
  \citenamefont {Abreu~Araujo}, \citenamefont {Riou}, \citenamefont {Torrejon},
  \citenamefont {Ravelosona}, \citenamefont {Kang}, \citenamefont {Zhao},
  \citenamefont {Grollier},\ and\ \citenamefont
  {Querlioz}}]{chen2022forecasting}%
  \BibitemOpen
  \bibfield  {author} {\bibinfo {author} {\bibfnamefont {X.}~\bibnamefont
  {Chen}}, \bibinfo {author} {\bibfnamefont {F.}~\bibnamefont {Abreu~Araujo}},
  \bibinfo {author} {\bibfnamefont {M.}~\bibnamefont {Riou}}, \bibinfo {author}
  {\bibfnamefont {J.}~\bibnamefont {Torrejon}}, \bibinfo {author}
  {\bibfnamefont {D.}~\bibnamefont {Ravelosona}}, \bibinfo {author}
  {\bibfnamefont {W.}~\bibnamefont {Kang}}, \bibinfo {author} {\bibfnamefont
  {W.}~\bibnamefont {Zhao}}, \bibinfo {author} {\bibfnamefont {J.}~\bibnamefont
  {Grollier}}, \ and\ \bibinfo {author} {\bibfnamefont {D.}~\bibnamefont
  {Querlioz}},\ }\href {\doibase 10.1038/s41467-022-28571-7} {\bibfield
  {journal} {\bibinfo  {journal} {Nature communications}\ }\textbf {\bibinfo
  {volume} {13}},\ \bibinfo {pages} {1} (\bibinfo {year} {2022})}\BibitemShut
  {NoStop}%
\bibitem [{\citenamefont {Thiele}(1973)}]{thiele1973steady}%
  \BibitemOpen
  \bibfield  {author} {\bibinfo {author} {\bibfnamefont {A.}~\bibnamefont
  {Thiele}},\ }\href {\doibase 10.1103/PhysRevLett.30.230} {\bibfield
  {journal} {\bibinfo  {journal} {Physical Review Letters}\ }\textbf {\bibinfo
  {volume} {30}},\ \bibinfo {pages} {230} (\bibinfo {year} {1973})}\BibitemShut
  {NoStop}%
\bibitem [{\citenamefont {Huber}(1982)}]{huber1982dynamics}%
  \BibitemOpen
  \bibfield  {author} {\bibinfo {author} {\bibfnamefont {D.}~\bibnamefont
  {Huber}},\ }\href {\doibase 10.1103/PhysRevB.26.3758} {\bibfield  {journal}
  {\bibinfo  {journal} {Physical Review B}\ }\textbf {\bibinfo {volume} {26}},\
  \bibinfo {pages} {3758} (\bibinfo {year} {1982})}\BibitemShut {NoStop}%
\bibitem [{\citenamefont {Abreu~Araujo}\ \emph
  {et~al.}(2022{\natexlab{b}})\citenamefont {Abreu~Araujo}, \citenamefont
  {Chopin},\ and\ \citenamefont
  {de~Wergifosse}}]{abreu_araujo_chopin_de_wergifosse_2022_splitting}%
  \BibitemOpen
  \bibfield  {author} {\bibinfo {author} {\bibfnamefont {F.}~\bibnamefont
  {Abreu~Araujo}}, \bibinfo {author} {\bibfnamefont {C.}~\bibnamefont
  {Chopin}}, \ and\ \bibinfo {author} {\bibfnamefont {S.}~\bibnamefont
  {de~Wergifosse}},\ }\href {\doibase 10.1038/s41598-022-14574-3} {\bibfield
  {journal} {\bibinfo  {journal} {Scientific Reports}\ }\textbf {\bibinfo
  {volume} {12}} (\bibinfo {year} {2022}{\natexlab{b}}),\
  10.1038/s41598-022-14574-3}\BibitemShut {NoStop}%
\bibitem [{\citenamefont {Guslienko}\ \emph {et~al.}(2014)\citenamefont
  {Guslienko}, \citenamefont {Sukhostavets},\ and\ \citenamefont
  {Berkov}}]{guslienko2014nonlinear}%
  \BibitemOpen
  \bibfield  {author} {\bibinfo {author} {\bibfnamefont {K.~Y.}\ \bibnamefont
  {Guslienko}}, \bibinfo {author} {\bibfnamefont {O.~V.}\ \bibnamefont
  {Sukhostavets}}, \ and\ \bibinfo {author} {\bibfnamefont {D.~V.}\
  \bibnamefont {Berkov}},\ }\href {\doibase 10.1186/1556-276X-9-386} {\bibfield
   {journal} {\bibinfo  {journal} {Nanoscale research letters}\ }\textbf
  {\bibinfo {volume} {9}},\ \bibinfo {pages} {1} (\bibinfo {year}
  {2014})}\BibitemShut {NoStop}%
\bibitem [{\citenamefont {Tanaka}\ \emph {et~al.}(2019)\citenamefont {Tanaka},
  \citenamefont {Yamane}, \citenamefont {H{\'e}roux}, \citenamefont {Nakane},
  \citenamefont {Kanazawa}, \citenamefont {Takeda}, \citenamefont {Numata},
  \citenamefont {Nakano},\ and\ \citenamefont {Hirose}}]{tanaka2019recent}%
  \BibitemOpen
  \bibfield  {author} {\bibinfo {author} {\bibfnamefont {G.}~\bibnamefont
  {Tanaka}}, \bibinfo {author} {\bibfnamefont {T.}~\bibnamefont {Yamane}},
  \bibinfo {author} {\bibfnamefont {J.~B.}\ \bibnamefont {H{\'e}roux}},
  \bibinfo {author} {\bibfnamefont {R.}~\bibnamefont {Nakane}}, \bibinfo
  {author} {\bibfnamefont {N.}~\bibnamefont {Kanazawa}}, \bibinfo {author}
  {\bibfnamefont {S.}~\bibnamefont {Takeda}}, \bibinfo {author} {\bibfnamefont
  {H.}~\bibnamefont {Numata}}, \bibinfo {author} {\bibfnamefont
  {D.}~\bibnamefont {Nakano}}, \ and\ \bibinfo {author} {\bibfnamefont
  {A.}~\bibnamefont {Hirose}},\ }\href@noop {} {\bibfield  {journal} {\bibinfo
  {journal} {Neural Networks}\ }\textbf {\bibinfo {volume} {115}},\ \bibinfo
  {pages} {100} (\bibinfo {year} {2019})}\BibitemShut {NoStop}%
\bibitem [{\citenamefont {Deng}(2012)}]{deng2012mnist}%
  \BibitemOpen
  \bibfield  {author} {\bibinfo {author} {\bibfnamefont {L.}~\bibnamefont
  {Deng}},\ }\href@noop {} {\bibfield  {journal} {\bibinfo  {journal} {IEEE
  Signal Processing Magazine}\ }\textbf {\bibinfo {volume} {29}},\ \bibinfo
  {pages} {141} (\bibinfo {year} {2012})}\BibitemShut {NoStop}%
\bibitem [{\citenamefont {Nair}\ and\ \citenamefont
  {Hinton}(2010)}]{nair2010rectified}%
  \BibitemOpen
  \bibfield  {author} {\bibinfo {author} {\bibfnamefont {V.}~\bibnamefont
  {Nair}}\ and\ \bibinfo {author} {\bibfnamefont {G.~E.}\ \bibnamefont
  {Hinton}},\ }in\ \href@noop {} {\emph {\bibinfo {booktitle} {Proceedings of
  the 27th international conference on machine learning (ICML-10)}}}\ (\bibinfo
  {year} {2010})\ pp.\ \bibinfo {pages} {807--814}\BibitemShut {NoStop}%
\bibitem [{\citenamefont {Narayan}(1997)}]{narayan1997generalized}%
  \BibitemOpen
  \bibfield  {author} {\bibinfo {author} {\bibfnamefont {S.}~\bibnamefont
  {Narayan}},\ }\href@noop {} {\bibfield  {journal} {\bibinfo  {journal}
  {Information sciences}\ }\textbf {\bibinfo {volume} {99}},\ \bibinfo {pages}
  {69} (\bibinfo {year} {1997})}\BibitemShut {NoStop}%
\bibitem [{\citenamefont {Schrauwen}\ \emph {et~al.}(2007)\citenamefont
  {Schrauwen}, \citenamefont {Verstraeten},\ and\ \citenamefont
  {Van~Campenhout}}]{schrauwen2007overview}%
  \BibitemOpen
  \bibfield  {author} {\bibinfo {author} {\bibfnamefont {B.}~\bibnamefont
  {Schrauwen}}, \bibinfo {author} {\bibfnamefont {D.}~\bibnamefont
  {Verstraeten}}, \ and\ \bibinfo {author} {\bibfnamefont {J.}~\bibnamefont
  {Van~Campenhout}},\ }in\ \href@noop {} {\emph {\bibinfo {booktitle}
  {Proceedings of the 15th european symposium on artificial neural networks. p.
  471-482 2007}}}\ (\bibinfo {year} {2007})\ pp.\ \bibinfo {pages}
  {471--482}\BibitemShut {NoStop}%
\bibitem [{\citenamefont {Furuta}\ \emph {et~al.}(2018)\citenamefont {Furuta},
  \citenamefont {Fujii}, \citenamefont {Nakajima}, \citenamefont {Tsunegi},
  \citenamefont {Kubota}, \citenamefont {Suzuki},\ and\ \citenamefont
  {Miwa}}]{furuta2018macromagnetic}%
  \BibitemOpen
  \bibfield  {author} {\bibinfo {author} {\bibfnamefont {T.}~\bibnamefont
  {Furuta}}, \bibinfo {author} {\bibfnamefont {K.}~\bibnamefont {Fujii}},
  \bibinfo {author} {\bibfnamefont {K.}~\bibnamefont {Nakajima}}, \bibinfo
  {author} {\bibfnamefont {S.}~\bibnamefont {Tsunegi}}, \bibinfo {author}
  {\bibfnamefont {H.}~\bibnamefont {Kubota}}, \bibinfo {author} {\bibfnamefont
  {Y.}~\bibnamefont {Suzuki}}, \ and\ \bibinfo {author} {\bibfnamefont
  {S.}~\bibnamefont {Miwa}},\ }\href@noop {} {\bibfield  {journal} {\bibinfo
  {journal} {Physical Review Applied}\ }\textbf {\bibinfo {volume} {10}},\
  \bibinfo {pages} {034063} (\bibinfo {year} {2018})}\BibitemShut {NoStop}%
\bibitem [{\citenamefont {Wold}\ \emph {et~al.}(1987)\citenamefont {Wold},
  \citenamefont {Esbensen},\ and\ \citenamefont {Geladi}}]{wold1987principal}%
  \BibitemOpen
  \bibfield  {author} {\bibinfo {author} {\bibfnamefont {S.}~\bibnamefont
  {Wold}}, \bibinfo {author} {\bibfnamefont {K.}~\bibnamefont {Esbensen}}, \
  and\ \bibinfo {author} {\bibfnamefont {P.}~\bibnamefont {Geladi}},\
  }\href@noop {} {\bibfield  {journal} {\bibinfo  {journal} {Chemometrics and
  intelligent laboratory systems}\ }\textbf {\bibinfo {volume} {2}},\ \bibinfo
  {pages} {37} (\bibinfo {year} {1987})}\BibitemShut {NoStop}%
\bibitem [{\citenamefont {Abdi}\ \emph {et~al.}(2013)\citenamefont {Abdi},
  \citenamefont {Williams}, \citenamefont {Reisfeld},\ and\ \citenamefont
  {Mayeno}}]{abdi2013computational}%
  \BibitemOpen
  \bibfield  {author} {\bibinfo {author} {\bibfnamefont {H.}~\bibnamefont
  {Abdi}}, \bibinfo {author} {\bibfnamefont {L.}~\bibnamefont {Williams}},
  \bibinfo {author} {\bibfnamefont {B.}~\bibnamefont {Reisfeld}}, \ and\
  \bibinfo {author} {\bibfnamefont {A.}~\bibnamefont {Mayeno}},\ }\href@noop {}
  {\  (\bibinfo {year} {2013})}\BibitemShut {NoStop}%
\bibitem [{mni(2023)}]{mnist_benchmark}%
  \BibitemOpen
  \href {https://paperswithcode.com/sota/image-classification-on-mnist}
  {\enquote {\bibinfo {title} {Papers with code - mnist benchmark (image
  classification)},}\ } (\bibinfo {year} {2023})\BibitemShut {NoStop}%
\bibitem [{\citenamefont {An}\ \emph {et~al.}(2020)\citenamefont {An},
  \citenamefont {Lee}, \citenamefont {Park}, \citenamefont {Yang},\ and\
  \citenamefont {So}}]{an2020ensemble}%
  \BibitemOpen
  \bibfield  {author} {\bibinfo {author} {\bibfnamefont {S.}~\bibnamefont
  {An}}, \bibinfo {author} {\bibfnamefont {M.}~\bibnamefont {Lee}}, \bibinfo
  {author} {\bibfnamefont {S.}~\bibnamefont {Park}}, \bibinfo {author}
  {\bibfnamefont {H.}~\bibnamefont {Yang}}, \ and\ \bibinfo {author}
  {\bibfnamefont {J.}~\bibnamefont {So}},\ }\href@noop {} {\bibfield  {journal}
  {\bibinfo  {journal} {arXiv preprint arXiv:2008.10400}\ } (\bibinfo {year}
  {2020})}\BibitemShut {NoStop}%
\bibitem [{Note1()}]{Note1}%
  \BibitemOpen
  \bibinfo {note} {The training of a $5000$-nodes neural reservoir on the whole
  training set took in average $47.3$ seconds on a Dell Precision 5820 tower
  equipped with an Intel Core i9-10980XE processor ($18$ cores and up to $36$
  threads), and $64.2$ seconds on a MacBook Pro equipped with the M2 SoC (12
  cores CPU).}\BibitemShut {Stop}%
\bibitem [{\citenamefont {Courrieu}(2008)}]{courrieu2008fast}%
  \BibitemOpen
  \bibfield  {author} {\bibinfo {author} {\bibfnamefont {P.}~\bibnamefont
  {Courrieu}},\ }\href@noop {} {\bibfield  {journal} {\bibinfo  {journal}
  {arXiv preprint arXiv:0804.4809}\ } (\bibinfo {year} {2008})}\BibitemShut
  {NoStop}%
\bibitem [{\citenamefont {Barata}\ and\ \citenamefont
  {Hussein}(2012)}]{barata2012moore}%
  \BibitemOpen
  \bibfield  {author} {\bibinfo {author} {\bibfnamefont {J.~C.~A.}\
  \bibnamefont {Barata}}\ and\ \bibinfo {author} {\bibfnamefont {M.~S.}\
  \bibnamefont {Hussein}},\ }\href@noop {} {\bibfield  {journal} {\bibinfo
  {journal} {Brazilian Journal of Physics}\ }\textbf {\bibinfo {volume} {42}},\
  \bibinfo {pages} {146} (\bibinfo {year} {2012})}\BibitemShut {NoStop}%
\bibitem [{\citenamefont {Tu}\ \emph {et~al.}(2023)\citenamefont {Tu},
  \citenamefont {Zhang}, \citenamefont {Luo}, \citenamefont {Lv}, \citenamefont
  {Lei}, \citenamefont {Cai}, \citenamefont {Fang}, \citenamefont {Finocchio},
  \citenamefont {Bian}, \citenamefont {Li} \emph {et~al.}}]{tu2023neural}%
  \BibitemOpen
  \bibfield  {author} {\bibinfo {author} {\bibfnamefont {H.}~\bibnamefont
  {Tu}}, \bibinfo {author} {\bibfnamefont {L.}~\bibnamefont {Zhang}}, \bibinfo
  {author} {\bibfnamefont {Y.}~\bibnamefont {Luo}}, \bibinfo {author}
  {\bibfnamefont {W.}~\bibnamefont {Lv}}, \bibinfo {author} {\bibfnamefont
  {T.}~\bibnamefont {Lei}}, \bibinfo {author} {\bibfnamefont {J.}~\bibnamefont
  {Cai}}, \bibinfo {author} {\bibfnamefont {B.}~\bibnamefont {Fang}}, \bibinfo
  {author} {\bibfnamefont {G.}~\bibnamefont {Finocchio}}, \bibinfo {author}
  {\bibfnamefont {L.}~\bibnamefont {Bian}}, \bibinfo {author} {\bibfnamefont
  {S.}~\bibnamefont {Li}},  \emph {et~al.},\ }\href@noop {} {\bibfield
  {journal} {\bibinfo  {journal} {Applied Physics Letters}\ }\textbf {\bibinfo
  {volume} {122}},\ \bibinfo {pages} {122402} (\bibinfo {year}
  {2023})}\BibitemShut {NoStop}%
\end{thebibliography}%

\end{document}